\documentclass[letterpaper,twocolumn,10pt]{article}
\usepackage{usenix2019_v3}

\usepackage{url}
\usepackage{comment}
\usepackage{graphicx}
\usepackage{subfig}
\usepackage{balance}
\usepackage{color}
\usepackage{hyperref}
\usepackage{url}

\usepackage{breakurl}
\usepackage{enumitem}

\usepackage{lipsum}
\usepackage{titlesec}

\titlespacing\section{0pt}{8pt plus 4pt minus 2pt}{6pt plus 2pt minus 2pt}
\titlespacing\subsection{0pt}{7pt plus 4pt minus 2pt}{6pt plus 2pt minus 2pt}
\titlespacing\subsubsection{0pt}{7pt plus 4pt minus 2pt}{6pt plus 2pt minus 2pt}

\usepackage{algorithm}
\usepackage{algorithmic}
\usepackage{amsmath}
\usepackage{amssymb}
\DeclareMathOperator*{\argmax}{arg\ max}

\usepackage{multirow}

\usepackage{authblk}

\newcommand{\gbdtrobusttimes}{1.25$\times$}
\newcommand{\gbdtbaselinetimes}{2.78$\times$}
\newcommand{\sklearnrfbaselinetimes}{3.52$\times$}
\newcommand{\sklearnrfrobusttimes}{1.7$\times$}
\newcommand{\twitterbaselinetimes}{10.6$\times$}

\begin{document}

\title{Cost-Aware Robust Tree Ensembles for Security Applications}

\renewcommand\Affilfont{\itshape}
{
	\author{Yizheng Chen}
	\author{Shiqi Wang}
	\author{Weifan Jiang}
	\author{Asaf Cidon}
	\author{Suman Jana}
	\affil{\vspace{-8pt}Columbia University}
}


\maketitle

\begin{abstract}

There are various costs for attackers to manipulate the features of security classifiers.
The costs are asymmetric across features and to the directions of changes,
which cannot be precisely captured by existing cost models based on $L_p$-norm robustness.  
In this paper, we utilize such domain knowledge to increase the attack cost of
evading classifiers, specifically, tree ensemble models that are widely used by
security tasks. We propose a new cost modeling method to capture
the feature manipulation cost as constraint, and then we integrate the cost-driven constraint
into the node construction process to train robust tree ensembles. During the training process,
we use the constraint to find data points that are likely to be perturbed given the feature manipulation cost,
and we use a new robust training algorithm to optimize the quality of the trees.
Our cost-aware training method can be applied to different types of tree ensembles, including
gradient boosted decision trees and random forest models. Using Twitter spam detection
as the case study, our evaluation results
show that we can increase the attack cost by \twitterbaselinetimes{} compared to the baseline.
Moreover, our robust training method using cost-driven constraint can achieve higher accuracy, lower false positive rate,
and stronger cost-aware robustness than the state-of-the-art training method using $L_\infty$-norm cost model.
Our code is available at \url{https://github.com/surrealyz/growtrees}.

\end{abstract}

\section{Introduction}

Many machine learning classifiers are used in security-critical settings where adversaries
actively try to evade them.
Unlike perturbing features (e.g., pixels, words) in other machine learning applications,
the attacker has different cost to manipulate different security features.
For example, to evade a spam filter, it is cheaper to purchase new domain names
than to rent new hosting servers~\cite{levchenko2011click}.
In addition, a feature may be expensive to increase, but easy to decrease.
For example, it is easier to remove a signature from a malware than to add one signed by
Microsoft to it~\cite{incer2018adversarially}.
We need cost modeling methods to capture such domain knowledge about feature manipulation cost,
and utilize the knowledge to increase the attack cost of evading security classifiers.

Since it is extremely hard, if not impossible, to be robust against
all attackers, we focus on making a classifier robust against
attackers bounded by the feature manipulation cost model.
To evaluate cost-aware robustness against unbounded attackers,
we measure the attack cost as the total cost
required to manipulate all features to evade the trained classifiers.
However, existing cost modeling based on
$L_p$-norm is not suitable for security applications, since it assumes uniform cost across different features
and symmetric cost to increase and decrease the features.
Moreover, many recent works focus on improving the robustness of neural network models~\cite{madry2017towards,sinha2018certifying,wong2018provable,wong2018scaling,wang2018mixtrain,gowal2018effectiveness,dvijotham2018training,mirman2018differentiable,li2018second,lecuyer2019certified,cohen2019certified},
whereas security applications widely use tree ensemble models such as random forest (RF)
and gradient boosted decision trees (GBDT) to detect malware~\cite{kwon2015dropper},
phishing~\cite{fette2007learning,ho2019detecting,cidon2019high},
and online fraud~\cite{nelms2016towards,rafique2016s,kharraz2018surveylance}, etc.
Despite their popularity, the robustness of these models, especially against a strong adversary is not very thoroughly studied~\cite{papernot2016transferability,kantchelian2016evasion,chen2019robust}.
The discrete structure of tree models brings new challenges to the robustness problem.
Training trees does not rely on gradient-guided optimization, 
but rather by enumerating potential splits to maximize the gain metric
(e.g., Information gain, Gini impurity reduction, or loss reduction).
It is intractable to enumerate all possible splits under attacks~\cite{chen2019robust}.

\begin{figure}[t!]
	\centering
	\includegraphics[width=\columnwidth]{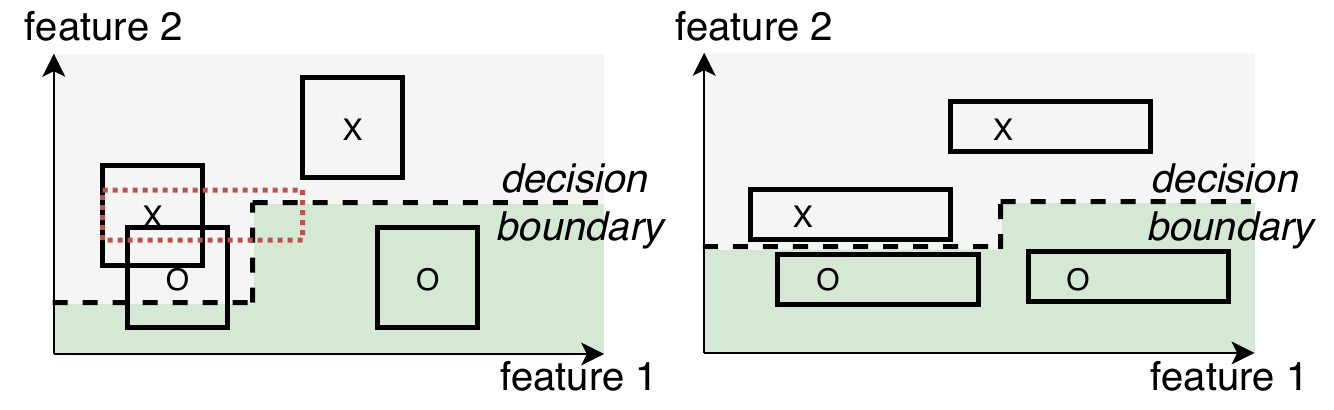}
	\caption{An example that we can obtain better model performance and cost-aware robustness if we use cost-driven constraints in robust training than $L_\infty$-norm bound. The dashed lines are the classification boundary. The left figure shows that robust training using $L_\infty$-norm bound (solid square box) achieves 75\% accuracy. Given the cost-aware perturbation (dashed red rectangular box), the model has only 50\% accuracy under attack. The right figure shows that using cost-driven constraint, we
    can achieve 100\% accuracy with and without attack.}
	\label{fig:motivation}
\end{figure}

Figure~\ref{fig:motivation} shows an example where we can obtain
better accuracy and stronger robustness
against realistic attackers, if we train a robust decision tree model
using knowledge about feature manipulation cost, instead of using the $L_\infty$-norm bound. In the left figure,
the square box denotes $L_\infty$-norm bound for four data
points with two feature dimensions.
We use the dashed line to denote the classification boundary of the robust split,
which can achieve 75\% accuracy, and 75\% robust accuracy against $L_\infty$-norm bounded attacks.
However, in practice, the classifier may not have 75\% robust accuracy against realistic attackers.
The dashed rectangular box for the cross on the left side
represents the realistic perturbation bound, that
it is easier to increase a data point than to decrease it
along feature 1, and it is harder to perturb feature 2 than feature 1.
Thus, the data point can be perturbed to evade the robust split,
and the actual robust accuracy is only 50\% against the realistic attack.
In comparison, if we can model the feature manipulation cost as constraints
for each data point's perturbation bound, we can learn the robust split in
the right figure and achieve 100\% accuracy and robust accuracy.

In this paper, we propose a systematic method to train cost-aware robust tree ensemble models for security,
by integrating domain knowledge of feature manipulation cost.
We first propose a cost modeling method that summarizes the domain knowledge
about features into cost-driven constraint, which represents
how bounded attackers can perturb every data point based
on the cost of manipulating different features.
Then, we integrate the constraint into the training process
as if an arbitrary attacker under the cost constraint is trying to maximally degrade the quality
of potential splits (Equation~\eqref{eq:max_uncertain} in Section~\ref{section:Optimization Problem}).
We propose an efficient robust training algorithm that solves
the maximization problem across different gain metrics and different types of models,
including random forest model and gradient boosted decision trees.
Lastly, we evaluate the adaptive attack cost against our robust training method, as a step towards understanding
robustness against attacks in the problem space~\cite{pierazzi2020intriguing}.
We propose an adaptive attack cost function to represent the total feature manipulation cost (Section~\ref{subsec:Adaptive Attacker}), as the minimization objective of the strongest whitebox attacker against tree ensembles (the Mixed Integer Linear Program attacker).
The attack objective specifically targets the cost-driven constraint, such that the attacker minimizes the total cost of perturbing different features.

\begin{figure}[t!]
	\centering
	\includegraphics[width=0.9\columnwidth]{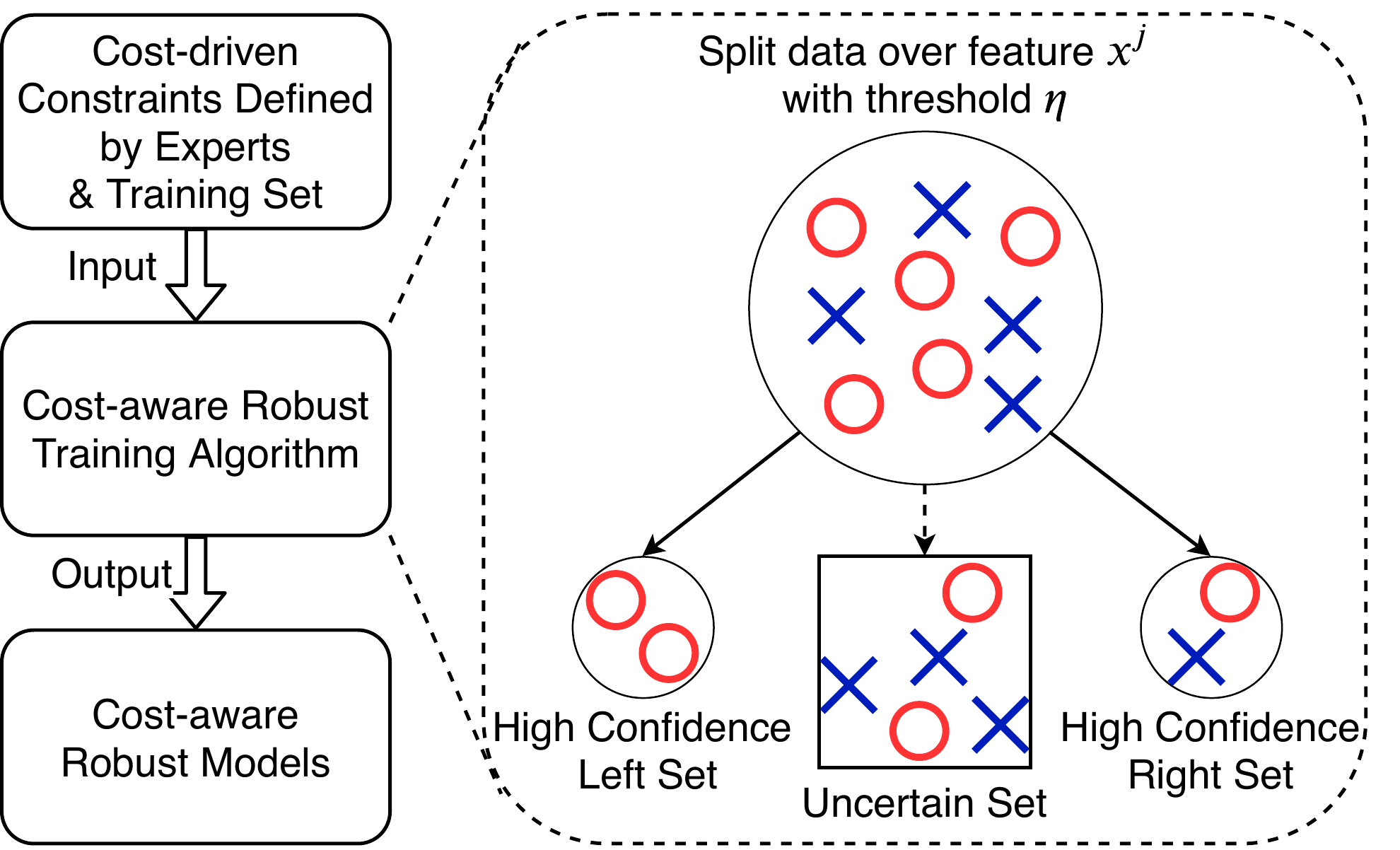}
	\caption{An overview of cost-aware robust tree ensemble training process.
		Our robust training algorithm incorporates the cost-driven constraint while constructing nodes.
		The constraint gives the set of data points that can potentially cross the split threshold $\eta$
		given domain knowledge about the $j$-th feature, i.e. uncertain set.}
	\label{fig:overview}
\end{figure}

Our robust training method incorporates the cost-driven constraint into the
node construction process of growing trees, as shown in Figure~\ref{fig:overview}.
When any potential split $x^j < \eta$ (on the $j$-th feature) is being considered,
due to the constraint, there is a range of possible values
a data point can be changed into for that feature (formally defined in Section~\ref{sec:Constraint Definition}).
Thus, data points close to the splitting
threshold $\eta$ can potentially cross the threshold. For example, on a low cost feature,
many data points can be easily perturbed to either the left child or the right child. Therefore,
the constraint gives us a set of uncertain data points that can degrade the quality of
the split, as well as high confidence data points that cannot be moved from the two children nodes.
We need to quantify the worst quality of the split under the constraint, in order
to compute the gain metric. To efficiently solve this, we propose a robust training algorithm that
iteratively assigns training data points to whichever side of the split with the worse gain,
regardless of the choice of the gain function and the type of tree ensemble model.
As an example, we can categorize every feature
into negligible, low, medium, or high cost to be increased and decreased by the attacker.
Then, we use a high-dimensional box as the constraint.
Essentially, the constraint gives the bounded attacker a larger increase (decrease) budget
for features that are easier to increase (decrease), and smaller budget for more costly features.
The cost-driven constraint helps the model learn robustness that can maximize the cost of evasion for the attacker.
We have implemented our training method in the state-of-the-art tree ensemble learning libraries:
gradient boosted decision trees in XGBoost~\cite{chen2016xgboost}
and random forest in scikit-learn~\cite{sklearn}.

We first evaluate the performance of our core training technique without the cost-driven constraint,
against regular training method as well as the state-of-the-art robust tree ensemble training method from Chen et al.~\cite{chen2019robust}.
In the gradient boosted decision trees evaluation, we reproduce existing results
to compare models over 4 benchmark datasets with the goal of improving robustness against attackers
bounded by $L_\infty$-norm (Section~\ref{sec:Training Algorithm Evaluation}).
Using the same settings of number of trees and maximal depth hyperparameters from existing work, our robust training algorithm achieves
on average \gbdtbaselinetimes{} and \gbdtrobusttimes{} improvement over regular training and state-of-the-art
robust training algorithm~\cite{chen2019robust}, respectively,
in the minimal $L_\infty$ distance required to evade the model.
In addition, we show that our algorithm provides better solutions
to the optimization problem
than the state-of-the-art~\cite{chen2019robust} in 93\% of the cases on average (Section~\ref{section:heuristic_benefit}).

In the random forest models evaluation, we have implemented Chen's algorithm
in scikit-learn since it was only available in XGBoost.
We first train 120 models in total to perform grid search over
number of trees and maximal depth hyperparameters.
Then, we choose the hyperparameters with the best validation accuracy
for each training algorithm and compare their robustness against
the strongest whitebox attack. 
On average over the four benchmarking datasets,
we achieve \sklearnrfbaselinetimes{} and \sklearnrfrobusttimes{}
robustness improvement in the minimal $L_\infty$
evasion distance compared to the baseline and Chen's algorithm, repsectively.
This shows that our core training technique
alone has made significant improvements to solve the robust optimization problem. 

Next, we evaluate the cost-aware robust training method for security,
using Twitter spam URL detection as a case study. We reimplement the feature extraction
over the dataset from~\cite{kwon2017domain} to detect malicious URLs
posted by Twitter spammers.
The features capture that attackers
reuse hosting infrastructure resources, use long redirection chains across different
geographical locations, and prefer flexibility of deploying different URLs.
Based on domain knowledge,
we specify four families of cost-driven constraints to train 19 different robust models,
with key results summarized as follows.
First, compared to regular training, our best model increases the cost-aware
robustness by \twitterbaselinetimes{}.
Second, our robust training method using cost-driven 
constraints can achieve higher accuracy, lower false positive rate,
and stronger cost-aware robustness than $L_\infty$-norm cost model
from Chen's algorithm~\cite{chen2019robust}.
Third, specifying larger perturbation range in the cost-driven constraint
generally decreases accuracy and increases false positive rate;
however, it does not necessarily increase the obtained robustness.
We need to perform hyperparameter tuning to find the best cost
model that balances accuracy and robustness.
Lastly, by training cost-aware robustness, we can also increase the model's robustness
against $L_1$ and $L_2$ based MILP attacks~\cite{kantchelian2016evasion}.


Our contributions are summarized as the following:
\begin{itemize}
\item We propose a new cost modeling method to translate domain knowledge about features into
cost-driven constraint. Using the constraint, we can train models to utilize domain knowledge outside the training data.

\item We propose a new robust training algorithm to train cost-aware robust tree ensembles for security,
by integrating the cost constraint. Our algorithm can be applied to both gradient boosted decision trees in XGBoost and random forest model in scikit-learn.




\item We use Twitter spam detection as the security application to train cost-aware robust tree ensemble models. Compared to regular training, our best model increases the attack cost to evade the model by \twitterbaselinetimes{}.
\end{itemize}

\section{Background and Related Work}
\subsection{Tree Ensembles}

A decision tree model guides the prediction path from the root to
a leaf node containing the predicted value, where each internal node
holds a predicate over some feature values.
An ensemble of trees consists of multiple decision trees, which aggregates the
predictions from individual trees. Popular aggregation functions include the
average (random forest) and the sum (gradient boosted decision tree) of
the prediction values from each decision tree.


\subsubsection{Notations}

We use the following notations for the tree ensemble in this paper.
The training dataset $D$ has $N$ data points with $d$ features
$D = \{(x_i, y_i)| i = 1, 2, ..., N\} (x_i \in \mathbb{R}^d, y \in \mathbb{R})$.
Each input $x_i$ can be written as a $d$-dimensional vector,
$x_i = [x_i^1, x_i^2, ..., x_i^d]$.
A predicate $p$ is in the form\footnote{Oblique trees which use multiple feature values in a predicate is rarely used in an ensemble due to high construction costs~\cite{norouzi2015efficient}.}
of $x^j < \eta$, which evaluates the $j$-th feature
$x^j$ against the split threshold $\eta$. Specifically, for the i-th training data,
the predicate checks whether $x_i^j < \eta$.
If $p = true$, the decision tree guides the prediction path to the left child,
otherwise to the right child. This process repeats until
$x_i$ reaches a leaf.
We use a function $f$ to denote a decision tree, which gives a real-valued
output for the input data point $x$ with the true label $y$.
For classification trees, $f(x)$ represents the predicted probability for
the true label $y$.




The most common decision tree learning algorithms use a greedy strategy
to construct the nodes from the root to the leaves, e.g., notably
CART~\cite{breiman1984classification}, ID3~\cite{quinlan1986induction}, and C4.5~\cite{quinlan1993c}.
The algorithm greedily picks the best feature $j^*$ and the
best split value $\eta^*$ for each node, which partitions the data points that
reach the current node ($I$) to the left child ($I_L$) and the right child ($I_R$),
i.e., $I = I_L \cup I_R$. The training algorithm optimizes the following objective using a scoring function to maximize the \emph{gain} of the split:


\begin{equation}
\label{eq:maxgain}
j^*, \eta^* = \argmax_{j, \eta} Gain(I_L, L_R) = \argmax_{j, \eta} (s(I) - s(I_L, I_R))
\end{equation}

In Equation~\eqref{eq:maxgain}, $s$ denotes a scoring function.
For example, we can use Shannon entropy,
Gini impurity, or any general loss function.
Splitting a node changes the score from $s(I)$ to $s(I_L, I_R)$.
For example, using the Gini impurity, we have $Gain(I_L, L_R) = Gini(I) - Gini(I_L, I_R)$.
A common strategy to solve Equation~\eqref{eq:maxgain} is
to enumerate all the features with all the possible split points to find the
maximum gain.
Starting from the root node, the learning algorithm chooses the best feature split with the maximum gain,
and then recursively constructs the children
nodes in the same way, until the score does not improve
or some pre-determined threshold (e.g., maximum depth) is reached.

A tree ensemble uses the weighted sum of prediction values from $K$ decision trees,
where $K$ is a parameter specified by the user. Each decision tree
can be represented as a function $f_t$.
Then, the ensemble predicts the output $\hat{y}$ as follows.

\begin{equation}
\hat{y} = \phi(x) = a*\sum_{t=1}^{K}{f_t(x)}
\end{equation}


Ensemble methods use bagging~\cite{breiman1996bagging} or
boosting~\cite{schapire1990strength,freund1995boosting,freund1997decision} to
grow the decision trees.
Random forest and gradient boosted decision tree (GBDT) are the most widely used tree ensembles. The random forest model uses $a=\frac{1}{K}$, and the GBDT
model set $a=1$. They use different methods to grow trees in parallel or sequentially, which we describe next.

\subsubsection{Random Forest}

A random forest model uses bagging~\cite{breiman1996bagging} to grow the trees
in parallel. Bagging, i.e., bootstrap aggregation, uses a random subset of the
training data and a random subset of features to train individual learners.
For each decision tree $f_t$, we first randomly sample $N'$ data points from $D$ to obtain the training dataset $D_t = \{(x_i, y_i)\}$, where $|D_t| = N'$ and
$N' \leq N$.
Then, at every step of the training algorithm that solves
Equation~\eqref{eq:maxgain}, we randomly select $d'$ features in $I$
to find the optimal split, where $d' \leq d$. The feature sampling is
repeated until we finish growing the decision tree. The training data and
feature sampling helps avoid overfitting of the model.


Random forest model has been used for various security applications,
e.g., detecting malware distribution~\cite{kwon2015dropper},
malicious autonomous system~\cite{konte2015aswatch},
social engineering~\cite{nelms2016towards},
phishing emails~\cite{fette2007learning,ho2019detecting,cidon2019high},
advertising resources for ad blocker~\cite{iqbal2020adgraph},
and online scams~\cite{rafique2016s,kharraz2018surveylance}, etc.
In some cases, researchers have analyzed
the performance of the model (e.g., ROC curve) given different subsets of
the features to reason about the predictive power of feature categories.

\subsubsection{Gradient Boosted Decision Tree}


Gradient boosted decision tree (GBDT) model uses boosting~\cite{schapire1990strength,freund1995boosting,freund1997decision} to grow the trees sequentially. Boosting iteratively train the learners, improving the new learner's performance by focusing
on data that were misclassified by existing learners. Gradient boosting
generalizes the boosting method to use an arbitrarily differentiable loss
function.

In this paper, we focus on the state-of-the-art GBDT training system
XGBoost~\cite{chen2016xgboost}.
When growing a new tree ($f_t$), all previous trees ($f_1$, $f_2$, ..., $f_{t-1}$) are fixed. Using $\hat{y}^{(t)}$ to denote the predicted value at the t-th iteration of adding trees, XGBoost minimizes the regularized loss $\mathcal{L}^{(t)}$ for the entire ensemble, as the scoring function in Equation~\eqref{eq:maxgain}.

\begin{equation}
\begin{aligned}
\mathcal{L}^{(t)} =& \sum_{i=1}^{n}{l(y_i, \hat{y}^{(t)}}) + \sum_{i=1}^{t}\Omega(f_i) \\
\end{aligned}
\end{equation}

In the equation, $l$ is an arbitrary loss function, e.g., cross entropy;
and $\Omega(f_i)$ is the regularization term, which captures the complexity of the i-th tree,
and encourages simpler trees to avoid overfitting.
Using a special regularization term,
XGBoost has a closed form solution to calculate the optimal gain
of the corresponding best structure of the tree, given a split $I_L$ and $I_R$ as the following.

\begin{equation}
\label{eq:xgboostgain}
\begin{aligned}
& Gain(I_L, L_R) \\
& = \frac{1}{2} \left[
 \frac{(\sum_{i\in I_L}g_i)^2}{\sum_{i\in I_L}h_i + \lambda}
+ \frac{(\sum_{i\in I_R}g_i)^2}{\sum_{i\in I_R}h_i + \lambda}
- \frac{(\sum_{i\in I}g_i)^2}{\sum_{i\in I}h_i + \lambda}
\right] - \gamma
\end{aligned}
\end{equation}

In particular, $g_i = {\partial_{\hat{y}^{(t-1)}}}(l(y_i, \hat{y}^{(t-1)})$ and
$h_i = {\partial_{\hat{y}^{(t-1)}}^2}(l(y_i, \hat{y}^{(t-1)})$ are the first order and second order gradients
for the $i$-th data point. In addition, $\gamma$ and $\lambda$ are hyperparameters related to the regularization term.

Boosting makes the newer tree dependent on previously grown trees.
Previously, random forest was considered to generalize better than
gradient boosting, since boosting alone could overfit the training
data without tree pruning, whereas bagging avoids that.
The regularization term introduced by xgboost significantly improves
the generalization of GBDT.

\subsection{Evading Tree Ensembles}
\label{sec:Evading Tree Ensembles}



There are several attacks against ensemble trees.
Among the blackbox attacks, Cheng et al.'s attack~\cite{cheng2018query}
has been demonstrated to work on ensemble trees. The attack minimizes the distance between
a benign example and the decision boundary, using a zeroth order optimization algorithm
with the randomized gradient-free method.
Papernot et al.~\cite{papernot2016transferability} proposed a whitebox attack based on heuristics. The attack searches for 
leaves with different classes within the neighborhood of the targeted leaf of the benign example,
to find a small perturbation that results in a wrong prediction.
In this paper, we evaluate the robustness of a tree ensemble by analyzing the potential evasion
caused by the strongest whitebox adversary, the Mixed Integer Linear Program (MILP) attacker.
The adversary has whitebox access to the model structure,
model parameters and the prediction score.
The attack finds the exact minimal evasion distance to the model if an adversarial example exists.

\noindent\textbf{Strongest whitebox attack: MILP Attack.} Kantchelian et al.~\cite{kantchelian2016evasion}
have proposed to attack tree ensembles by constructing a mixed integer linear program,
where the variables of the program are
nodes of the trees, the objective is to minimize a distance (e.g., $L_p$ norm distance) between
the evasive example and the attacked data point, and the constraints of the program
are based on the model structure. The constraints include model mislabel requirement,
logical consistency among leaves and predicates. Using a solver, the MILP attack can find
adversarial example with the \emph{minimal evasion distance}.
Otherwise, if the solver says the program is infeasible,
there truly does not exist an adversarial example by perturbing the attacked data point.
Since the attack is based on a linear program, we can use it to minimize any objective in
the linear form.

\noindent\textbf{Adversarial training limitation.}
The MILP attack cannot be efficiently used for adversarial training,
e.g., it can take up to an hour to generate one adversarial example~\cite{chen2019robustness}
depending on the model size. Therefore, we integrate the cost-driven constraint
into the training process directly to let the model learn knowledge about features.
Moreover, Kantchelian et al.~\cite{kantchelian2016evasion}
demonstrated that adversarial training using a fast attack algroithm
that hardens $L_0$ evasion distance makes
the model weaker against $L_1$ and $L_2$ based attacks.
Our results demonstrate that by training cost-aware robustness,
we can also enhance the model's robustness against $L_1$ and $L_2$
based attacks.




\subsection{Related Work}
\label{subsec:Related Work}

From the defense side, existing robust trees training algorithms~\cite{andriushchenko2019provably,chen2019robust,wang2020ell_p}
focus on defending against $L_p$-norm bounded attackers,
which may not capture the attackers' capabilities in many applications.
Incer et al.~\cite{incer2018adversarially} train monotonic classifiers with the assumption that
the cost of decreasing some feature values is much higher compared to increasing them,
such that attackers cannot evade by increasing feature values.
In comparison, we model difficulties in both increasing and decreasing feature values,
since decreasing security features can incur costs of decreased utility~\cite{chen2017practical,kharraz2019outguard}.
Zhang and Evans~\cite{zhang2018cost} are the first to train cost-sensitive robustness
with regard to classification output error costs, since
some errors have more catastrophic consequences than others~\cite{dreossi2018semantic}.
Their work models the cost of classifier's output, whereas we model the cost of perturbing
the input features to the classifier.

The work most related to ours is from Calzavara et al.~\cite{calzavara2019adversarial,calzavara2020treant}.
They proposed a threat model that attackers can use attack rules,
each associated with a cost, to exhaust an attack budget.
The attack rules have the advantage of accurately perturbing categorical features
by repeatedly corrupting a feature value based on a step size.
Their training algorithm indirectly computes a set of data points that can be
perturbed based on all the possible combinations of the rules,
which in general needs enumeration and incurs an expensive computation cost.
In comparison, we map each feature value to perturbed ranges, which
directly derives the set of data points that can cross the splitting threshold
at training time without additional computation cost.
Our threat model has the same expressiveness as their rule-based model. For example,
by specifying the same perturbation range for every feature, we
can capture attackers bounded by $L_\infty$-norn and $L_1$-norm distances.
We could also model attacks that change categorical features by
using conditioned cost constraint for each category.
In addition, we can easily incorporate
our cost-driven constraints on top of state-of-the-art algorithm~\cite{wang2020ell_p} to train for 
attack distance metrics with dependencies among feature, e.g., constrained $L_1$ and $L_2$ distances.



Researchers have also modeled the cost in the attack objective. Lowd and Meek~\cite{lowd2005adversarial} propose a linear attack cost function to model
the feature importance. It is a weighted sum of absolute feature differences
between the original data point and the evasive instance.
We design adaptive attack objectives to minimize linear cost functions in a similar
way, but we assign different weights to different feature change directions.
Kulynych et al.~\cite{kulynych2018evading} proposed to model attacker's capabilities
using a transformation graph, where each node is an input that the attacker can
craft, and each directed edge represents the cost of transforming the input.
Their framework can be used to find minimal-cost adversarial examples,
if there is detailed cost analysis available for the atomic transformations
represented by the edges.

Pierazzi et al.~\cite{pierazzi2020intriguing} proposed several problem-space
constraints to generate adversarial examples in the problem space,
e.g., actual Android malware. Their constraints are related to the cost factors
we discuss for perturbing features. In particular, a high cost feature perturbation
is more likely to violate their problem-space constraints, compared to a low cost change.
For example, changes that may affect the functionality of a malware can violate
the preserved semantics constraint, and attack actions that increases the
suspiciousness of the malicious activity could violate the plausibility constraint.
We could set a total cost budget for each category of the cost factors (similar to ~\cite{calzavara2019adversarial,calzavara2020treant}), to ensure that
problem-space constraints are not violated, which we leave as future work.



\section{Methodology}

In this section, we present our methodology to train robust tree ensembles
that utilize expert domain knowledge about features.
We will describe how to specify the attack cost-driven constraint
that captures the domain knowledge, our robust
training algorithm that use the constraint, and a new
adaptive attack objective to evaluate the robust model.

\subsection{Attack Cost-driven Constraint}


\subsubsection{Constraint Definition}
\label{sec:Constraint Definition}

We define the cost-driven constraint for each feature $x^j$ to be $C(x^j)$. It is a mapping from the interval $[0, 1]$,
containing normalized feature values, to a set in $[0, 1] \times [0, 1]$. For each concrete value of $x^j$,
$C(x^j)$ gives the valid feature manipulation interval for any bounded attacker according to the cost of changing the feature,
for \emph{all training data points.}

\begin{figure}[t!]
	\centering
	\includegraphics[width=0.9\columnwidth]{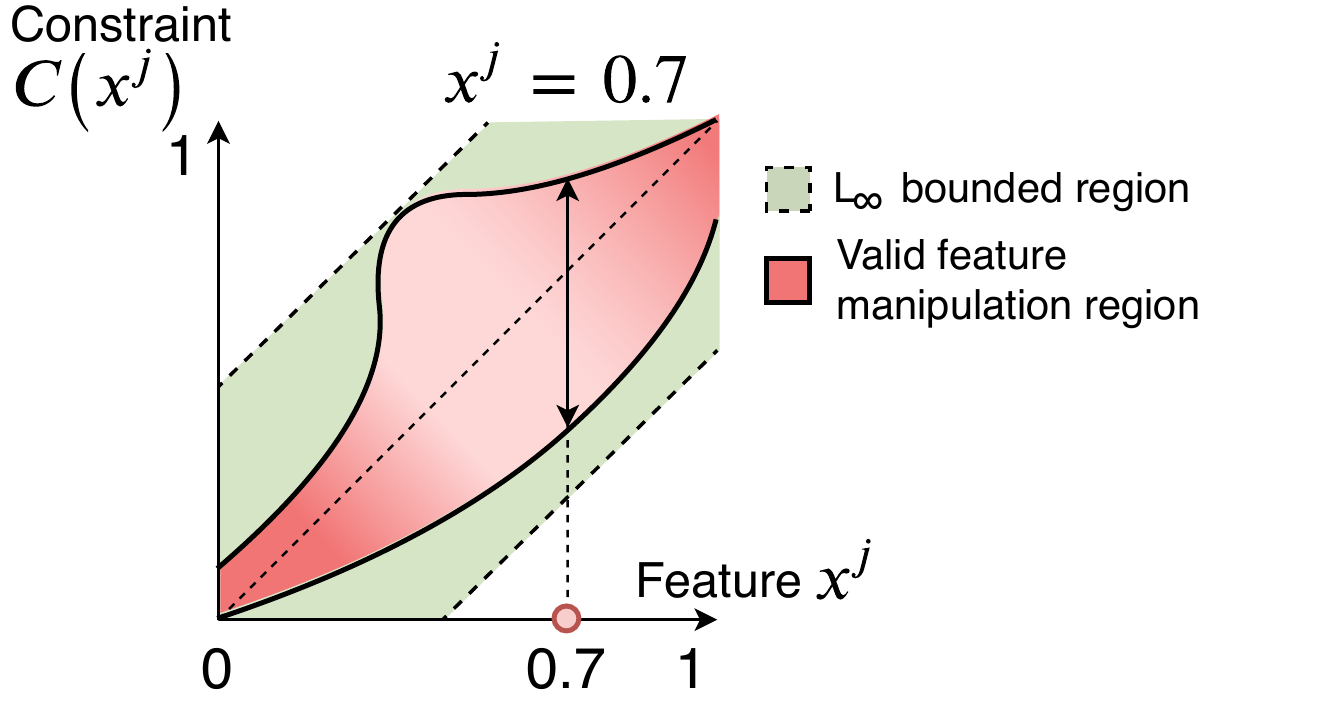}
	\caption{An example of cost-driven constraint for feature $x^j$. The red area represents valid feature manipulation region under the given cost-driven constraint $C(x^j)$ while the green area represents the common $L_\infty$-norm bounded region. Lighter red color means lower cost region, such that these feature values
		can be perturbed more by the attacker. The $L_\infty$ region is imprecise to capture the cost.}
	\label{fig:constraint}
\end{figure}

Figure~\ref{fig:constraint} shows two examples of cost-driven constraints.
We use the shaded region within the dashed lines to denote the constraint
when the attack cost is modeled by $L_\infty$-norm $\leq 0.5$ for feature $x^j$.
As $x^j$ takes different feature values between 0 and 1, $L_\infty$ cost model
states that the allowable perturbations for the feature are within
$[x^j-0.5, x^j+0.5]$. However, this region can be imprecise according
to the meaning of the feature. If the cost of changing feature $x^j$
is high when the value is close to 0 and 1, and relatively low in the middle,
we could have a valid feature manipulation region represented by the red area
enclosed in solid curves. When $x^j=0.7$ for one data point, the constraint says
that the cost is acceptable for the attacker to perturb $x^j$ between 0.45 and 0.90.
Lighter colored region allows larger perturbations than the darker colored region.
In general, the constraint can be anything
specified by the domain expert, by considering different cost factors.


\subsubsection{Cost Factors}

Different factors may affect the attack cost of feature perturbation,
and provide some general guidelines in ranking the cost across features
and their values.

\noindent\textbf{Economic.} The economic return on the attacker's investment is a major motivation
to whether they are willing to change some features. For example, to evade blocklist detection,
the attackers constantly register new domains and rent new servers. Registering new domains
is preferred since this costs less money than renting a new server.

\noindent\textbf{Functionality.}
Some features are related to malicious functionalities of the attack. For example,
the cryptojacking classifier~\cite{kharraz2019outguard} uses a feature that indicates
whether the attack website calls the CryptoNight hashing function to mine Monero coins.
Without function aliases, it is a high cost to remove the hash function since the website can no longer mine coins.

\noindent\textbf{Suspiciousness.}
If the attacker needs to generate a lot more malicious activities
to perturb features (e.g., sending more tweets than 99\% of users), this makes the attack easier to be
detected and has a cost.

\noindent\textbf{Monotonicity.}
In security applications, the cost to increase a feature
may be very different from decreasing it. For example, to evade malware detector
that uses ``static import'' features, it is easier to insert
redundant libraries than to remove useful ones~\cite{incer2018adversarially}.
Therefore, we need to specify the cost for both directions of the change.

\noindent\textbf{Attack seed.}
If the attack starts modifying features from a benign example (e.g., reverse mimicry attack),
the cost of changing features may be different from modifying features from a malicious data point.
Therefore, the seed sample can affect the cost.

\noindent\textbf{Ranking feature cost.}
Before specifying the constraints, we can roughly rank the cost
of manipulating different features and different values of the same feature.
All the cost factors can be translated to some return
over investment for attackers. Perturbing a feature could cost the attacker
more investment to set up the attack infrastructure,
purchase more compromised machines, or obtain some expensive benign samples.
On the other hand, feature perturbation could
reduce the revenue of malicious activities by eliminating certain functionalities,
or sacrificing some bots to be detected. From the perspective of both the investment
and the return, a domain expert can rank features by attack costs,
which is useful to construct the cost-driven constraint.
In addition, we can use
intervals or a continuous function of the value
to rank the cost of perturbing different values for the same feature.
Given the cost ranking of features, we provide two example constraints
below.

\subsubsection{Box Cost Constraint}
\label{sec:Box Cost constraint}


As an example, we describe how to specify the box constraint to capture
the domain knowledge about feature manipulation cost.
After analyzing the cost factors and ranking the feature manipulation cost,
we categorize attacker's cost of increasing and decreasing
the value of each feature into one of the four categories: negligible, low, medium, and high costs.
The categories are based on relative cost differences,
rather than absolute scale.

\textbf{Negligible cost.} There is negligible cost to perturb some features. For example,
in the code transformation attack against authorship attribution classifier, the attacker
can replace a for loop with a while loop without modifying any functionality of the code~\cite{quiring2019misleading}.
This changes the syntactic features of the classifier but incurs negligible costs.

\textbf{Low and medium cost.} Altering some features generates low or medium level
of costs by comparison.
For example, registering a new phishing domain name
is generally considered to be lower cost for the attacker than renting and
maintaining a new hosing server~\cite{levchenko2011click}.
Therefore, increasing domain name count features can be categorized as low cost,
whereas increasing IP address count features is medium cost.

\textbf{High cost.} If changing a feature significantly reduces the attack effectivenss,
or compromises the attacker, then it is a high cost feature.

\textbf{Box constraint.} After assigning different categories to increasing/decreasing features,
we can map the knowledge into a high dimensional
box as the following.

\begin{equation}
C(x^j) = [x^j-l_j, x^j+h_j],\ j= 1, 2, 3, ..., d
\end{equation}

It means that for the $j$-th feature, the constraint maps the feature
to the interval $[x^j-l_j, x^j+h_j]$ that represents the attacker's allowable
changes on the $j$-th feature value by decreasing or increasing it.
According to the category of cost for decreasing and increasing the $j$-th feature,
we can assign concrete values for $l_j$ and $h_j$.
These values can be hyperparameters for the training procedure.
Table~\ref{tab:cost_table} shows a mapping from the four categories to hyperparameters
$\alpha, \beta, \gamma$, and $\mu$, representing the percentage of change with regard to the
maximal value of the feature. A higher cost category should allow a smaller percentage of change
than a lower cost category, and thus, $\mu < \gamma < \beta < \alpha$.

\begin{table}[t!]
	\centering
	\small
	\begin{tabular}{l | r}
		\hline
		\textbf{Cost} & \textbf{Value for $l_j$, $h_j$} \\
		\hline
		\hline
		Negligible & $\alpha$ \\
		Low & $\beta$ \\
		Medium & $\gamma$ \\
		High & $\mu$ \\
		\hline
		Relationship & $\mu<\gamma<\beta<\alpha$\\
		\hline
		\hline
	\end{tabular}
	\caption{
		Feature manipulation cost categories based on domain knowledge.
		For each feature $j$, we categorize the cost of increasing and decreasing its values and assign the bound
		for the box constraint using variables $l_j$ and $h_j$.
		\label{tab:cost_table}}
\end{table}

For each dimension $j$ of every training data point $x_i$, the constraint says
that the attacker is allowed to perturb the value $x_i^j$ to $[x_i^j-l_j, x_i^j+h_j]$.
We will use this information to compute the gain of the split on $x^j$
in Equation~\eqref{eq:maxgain}. When comparing the quality of the splits,
the cost-driven constraint changes how the gain is evaluated, which we will
formalize in Equation~\eqref{eq:max_uncertain}.
In particular, if a split threshold is within the perturbation interval,
the training data point can be perturbed to cross the split threshold and
potentially evade the classifier, which degrades the gain of the split.

\subsubsection{Conditioned Cost Constraint}
As another example,
we can design the cost-driven constraint based on different conditions of the data point,
e.g., the attack seed factor.
In addition, the constraint can vary for different feature values. For example, we can design the constraint in Equation~\eqref{eq:general const} where $x_i^j$ denotes the $j$-th feature value of data point $x_i$.

\begin{equation}
C(x_i^j) = \begin{cases}
               0               & x_i\ \text{is benign} \\
               [x_i^j, 1]               & x_i\ \text{is mal, pred score > 0.9} \\
               [-0.1, 0.1]*x_i^j      & x_i\ \text{is mal, pred score <= 0.9} \\
               \end{cases}
\label{eq:general const}
\end{equation}
 
In this example, we give different constraints for benign and malicious data points for the $j$-th feature.
If a data point $x_i$ is benign, we assign a value zero, meaning that it is extremely hard for the
attacker to change the $j$-th feature value for a benign data point.
If the data point is malicious, we separate to two cases. When the prediction score
is higher than 0.9, we enforce that $x_i^j$ can only be increased.
On the other hand, when the prediction score is less than or equal to 0.9, we allow a relative 10\% change
for both increase and decrease directions, depending on the original value of $x_i^j$.

When evaluating the gain of the split in the training process,
we can use this constraint to derive the set of training data points under attack for every feature dimension $j$
and every split threshold $\eta$ as following. First, we calculate the prediction confidence
of a training data point by using the entire tree model. If the prediction score is larger than
0.9, we take every malicious data point with $x_i^j < \eta$. Otherwise, we take
all malicious data points with $x_i^j \in [\frac{1}{1.1}\eta, \frac{1}{0.9}\eta]$ to calculate
the reduced gain of the split.
We don't consider benign data points to be attacked in this case.

Our threat model has the same expressiveness as the rule-based model in~\cite{calzavara2020treant}. Our approach to use the cost driven constraint
directly maps each feature value to perturbed ranges, which
can be more easily integrated in the robust training algorithm compared to the rule-based
threat model.

\subsection{Robust Training}
\label{section:Methodology:Robust Training}


Given attack cost-driven constraints specified by domain experts, we propose
a new robust training algorithm that can integrate such information into the tree ensemble
models.

\subsubsection{Intuition}

\begin{figure}
	\centering
	\includegraphics[width=0.8\columnwidth]{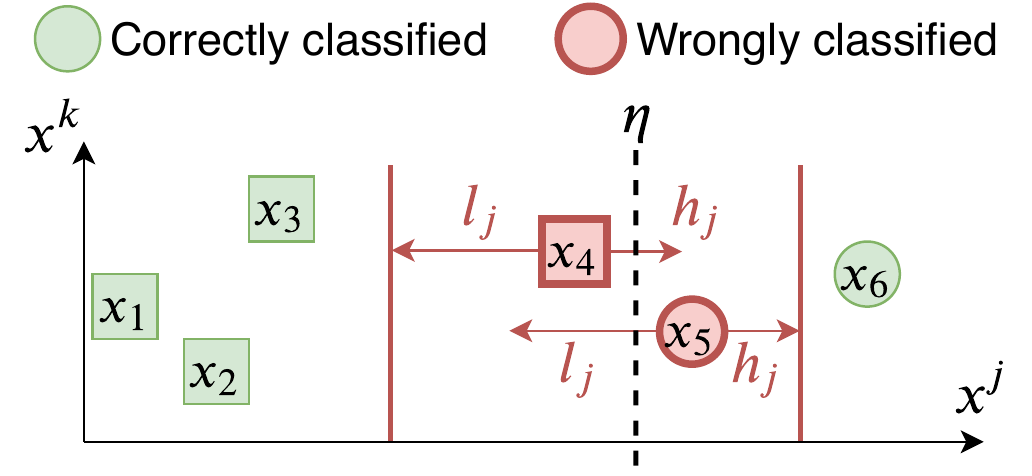}
	\includegraphics[width=0.8\columnwidth]{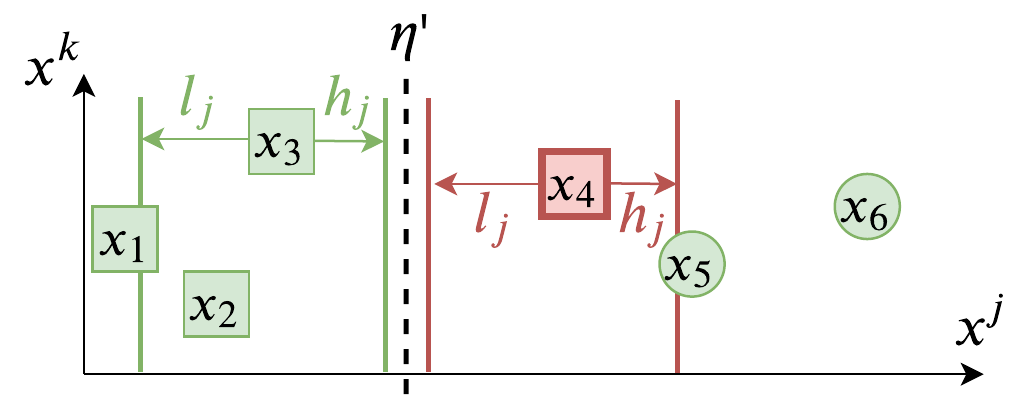}
	\caption{The intuition behind the attack cost-driven constraints for robust training, given six training points with two different classes (square and circle). It is easier to decrease $x^j$ than to increase it for the attacker. In the top figure, the split is 100\% accurate without attacks, but only 66.6\% accurate under attacks. The split in the bottom figure is always robust, but has a 83.3\% accuracy.
	}
	\label{fig:intuition}
\end{figure}

Using the box constraint as an example, we present the intuition of our robust training algorithm in Figure~\ref{fig:intuition}. The regular training algorithm of tree ensemble finds a non-robust split (top),
whereas our robust training algorithm can find a robust split (bottom) given the attack cost-driven constraint.
Specifically in the example, it is easier to decrease the feature $x^j$ than to increase it.
The cost constraint to increase (decrease) any $x^j$ is defined by $h_j$ ($l_j$).
Here $x_1,...,x_6$ are six training points with two different labels. The top of Figure~\ref{fig:intuition} shows that,
in regular training, the best split threshold $\eta$ over feature $x^j$ is
between $x_4$ and $x_5$, which perfectly separates the data points into left and right sets. However, given the attack cost to change feature $x^j$, $x_4$ and $x_5$ can both be easily perturbed by the adversary and cross the splitting threshold $\eta$. Therefore, the worst case accuracy under attacks is 66.6\%,
although the accuracy is 100\% without attacks. By integrating the
attack cost-driven constraints, we can choose a more robust split, as shown in
the bottom of Figure~\ref{fig:intuition}. Even though the attacker can
increase $x_3^j$ by up to $h_j$, and decrease $x_4^j$ by
up to $l_j$, the data points cannot cross the robust split threshold
$\eta'$. Therefore, the worst case accuracy under attacks is 83.3\%,
higher than that from the naive split.
As a tradeoff, $x_4$ is wrongly separated without attacks, which results in
83.3\% regular test accuracy  as well. As shown in the figure,
using a robust split can increase the minimal evasion distance for the attacker to cross
the split threshold.


\subsubsection{Optimization Problem}
\label{section:Optimization Problem}

In robust training, we want to maximize the gain computed from potential splits (feature $j$ and threshold $\eta$), 
given the domain knowledge about how robust a feature $x^j$ is.
We use $C$ to denote the attack cost-driven constraint.
Following Equation~\eqref{eq:maxgain}, we have the following:

\begin{equation}
\begin{aligned}
j^*, \eta^* & = \argmax_{j, \eta} Gain(I_L, L_R, C) \\
& = \argmax_{j, \eta} (s(I, C) - s(I_L, I_R, C)) \\
& = \argmax_{j, \eta} (s(I) - s(I_L, I_R, C))
\end{aligned}
\end{equation}

\textbf{Project constraint into set $\Delta I$.}
Since perturbing the
feature does not change the score $s(I)$ before the split ($s(I,C)$ is the same as $s(I)$), this only affects the score
$s(I_L, I_R, C)$ after the split, which cannot be efficiently computed.
Therefore, we project the second term as the worst case score
conditioned on some training data points $\Delta I$ being perturbed given the constraint function.
The perturbations degrade the quality of the split to two children sets $I'_L$ and $I'_R$
that are more impure or with higher loss values. To best utilize the feature manipulation cost knowledge,
we optimize for the maximal value of the score after the split, given different children sets $I'_L$ and $I'_R$
under the constraint. We then further categorize them into the high confidence points $I_{L_c}$ on the left side, $I_{R_c}$ on the right side, and low confidence points
$\Delta I_L$ and $\Delta I_R$:

\begin{equation}
\label{eq:max_uncertain}
\begin{aligned}
s(I_L, I_R, C) & = \max_{I'_L, I'_R, C} s(I'_L, I'_R) \\
& = \max_{\Delta I_L, \Delta I_R} s(I_{L_c} \cup \Delta I_L, I_{R_c} \cup \Delta I_R) \\
\end{aligned}
\end{equation}

\begin{figure}[t!]
	\centering
	\includegraphics[width=0.8\columnwidth]{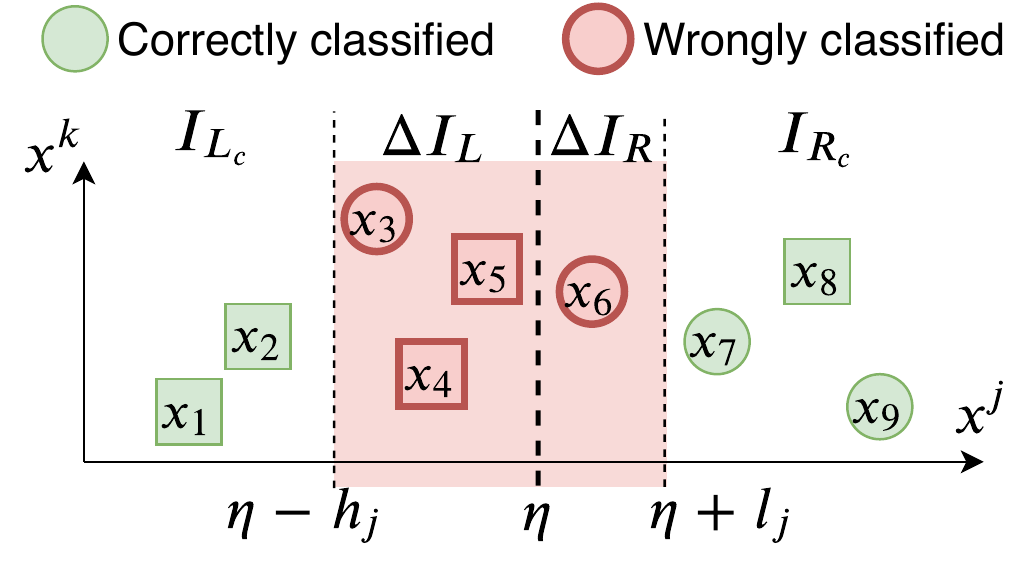}
	\caption{A simple example to illustrate the uncertain set $\Delta I= \Delta I_L \cup \Delta I_R=[x_3, x_4, x_5, x_6]$ within the robust region $[\eta-h_j, \eta+l_j]$ on feature $x^j$. The split threshold $\eta$ separates the data points into high confidence left set $I_{L_c}$ and high confidence right set $I_{R_c}$. Attackers can perturb the data points within the uncertain set $\Delta I_L \cup \Delta I_R$ to cross the split threshold, but not the high confidence data points.}
	\label{fig:uncertain}
	\vspace{-10pt}
\end{figure}

\textbf{Example.} Different constraint functions result in different $\Delta I$ set.
As an example, Figure~\ref{fig:uncertain} explains how we can map the box constraint for the $j$-th feature
to an uncertain set $\Delta I$ containing variables to be optimized.
We have nine data points numbered from 1 to 9,
i.e. $I = \{x_1, x_2, ..., x_9\}$, with two classes shaped in circles and squares.
The training process tries to put the splitting threshold $\eta$ between every two consecutive
data points, in order to find the split with maximum gain (Equation~\eqref{eq:maxgain}).
In Figure~\ref{fig:uncertain}, the split value under consideration is between data points $x_5$ and $x_6$.
The regular training process then computes the gain of the split based on
$I_L = \{x_1, x_2, x_3, x_4, x_5\}$ and $I_R = \{x_6, x_7, x_8, x_9\}$, using Equation~\ref{eq:maxgain}. 
In the robust training process, we first project the box constraint for the feature $j$
into the uncertain set $\Delta I = \{x_3, x_4, x_5, x_6\}$. Since the points on the left side of threshold $\eta$
can be increased by up to $h_j$, and points on the right side of $\eta$ can be decreased by up to $l_j$,
we get the shaded region of $[\eta-h_j, \eta+l_j]$ containing four data points that can be perturbed
to cross the splitting threshold $\eta$. Then, we need to maximize the score after split under the box constraint.
Each point in $\Delta I$ can be assigned to either the left side or the right side $\Delta I = \Delta I_L \cup \Delta I_R$,
with $2^{|\Delta I|}$ possible assignments. Finding the minimal gain assignment
is a combinatorial optimization problem, which needs to be repeatedly solved during the training process.
Therefore, we propose a new algorithm to efficiently solve Equation~\eqref{eq:max_uncertain}.

\subsubsection{Robust Training Algorithm}

\begin{algorithm}[t!]
\caption{Robust Training Algorithm}
\label{alg:robustsplitter}
\begin{flushleft}
\textbf{Input:} training set $D = \{(x_i, y_i)\}, |D|=N$
$(x_i \in \mathbb{R}^d, y \in \mathbb{R})$. \\
\textbf{Input:} data points of the current node
$I = \{(x_i, y_i)\}$, $|I| = m$.\\
\textbf{Input:} attack cost-driven constraint $C$. \\
\textbf{Input:} the score function $s$. \\
\textbf{Output:} the best split at the current node $j^*$, $\eta^*$. \\
\end{flushleft}
\begin{algorithmic}[1]
\STATE Initialize $Gain^* = 0; j^* = 0; \eta^* = 0$
\FOR{$j=1$ \TO $d$} 
	\STATE Sort $I = \{(x_i, y_i)\}$ along the j-th feature as $\{(x_{t_i}, y_{t_i})\}$
	\FOR{$t_i = t_1$ \TO $t_m$}
		\IF{$t_i = t_1$}
			\STATE $\eta \gets x_{t_1}^j$
		\ELSE
			\STATE $\eta \gets \frac{1}{2}(x_{t_i}+x_{t_{i-1}})$
		\ENDIF
		\STATE Project $C$ to the uncertain set $\Delta I$.
		\STATE $I_L = \{ (x_i, y_i) | x_i^j < \eta, x \notin \Delta I\}$
		\STATE $I_R = \{ (x_i, y_i) | x_i^j > \eta, x \notin \Delta I\}$
		\STATE /* Greedily put $(x_k, y_k)$ to whichever side that has a larger score to solve Equation~\eqref{eq:max_uncertain}. */
		\FOR{every $(x_k, y_k)$ in $\Delta I$}
			\STATE $ls = s(I_L\cup\{(x_k, y_k)\}, I_R)$
			\STATE $rs = s(I_L, I_R\cup\{(x_k, y_k)\})$
			\IF{$ls > rs$}
				\STATE $I_L =  I_L\cup\{(x_k, y_k)\}$
			\ELSE
				\STATE $I_R = I_R\cup\{(x_k, y_k)\}$ 
			\ENDIF
		\ENDFOR
		\STATE /* Find the maximal  gain. */
		\STATE $Gain(j, \eta, I) = s(I) - s(I_L, I_R)$
		\IF{$Gain(j, \eta, I) > Gain^*$}
			\STATE $j^* = j; \eta^* = \eta$
			\STATE $Gain^* = Gain(j, \eta, I)$
		\ENDIF
	\ENDFOR
\ENDFOR
\RETURN $j^*$, $\eta^*$
\end{algorithmic}
\end{algorithm}

We propose a new robust training algorithm to efficiently solve
the optimization problem in Equation~\eqref{eq:max_uncertain}.
Our algorithm works for different types of trees, including both classification and regression
trees, different ensembles such as gradient boosted decision trees and random forest,
and different splitting metrics used to compute the gain.

Algorithm~\ref{alg:robustsplitter} describes our robust training algorithm.
The algorithm provides the optimal splitting feature $j^*$ and the splitting threshold $\eta^*$
as output. The input includes the training dataset, the set of data points that reach
the current node $I = \{(x_i, y_i)\}$, the attack cost-driven constraint function, and a score function $s$.
Example score functions are the cross-entropy loss, Gini impurity, or
Shannon entropy.
From \texttt{Line 10} to \texttt{Line 28}, the algorithm does robust training, and the loops outside that are the  procedure used in regular training algorithm.
The algorithm marches through every feature dimension (the for loop at \texttt{Line 2}),
to compute the maximal score after the split given the feature manipulation cost knowledge,
for every possible split on that feature dimension. For each feature $j$, we first sort all the
data points along that dimension (\texttt{Line 3}). Then, we go through all the
sorted data points $(x_{t_i}, y_{t_i})$ to consider the gain of a potential split $x^j < \eta$ where $\eta$ is calculated from \texttt{Line 5} to \texttt{Line 9}.
Given the constraint function C, we project that to the uncertain set $\Delta I$
and initialize two more sets:
$I_L$ contains the data points that stay on the left side of the split,
and $I_R$ contains the data points that stay on the right (\texttt{Line 10 to 12}).
Next, from \texttt{Line 13} to \texttt{Line 22},
we go through every uncertain data point, and greedily put it to either $I_L$ or $I_R$,
whichever gives a larger score for the current split, to solve Equation~\eqref{eq:max_uncertain} under attacks. After that, we compute the gain under attacks
at \texttt{Line 24}, and update the optimal
split $j^*$, $\eta^*$ for the current node if the current gain is the largest (\texttt{Line 25 to 28}).
The algorithm eventually returns the optimal split ($j^*$, $\eta^*$) on \texttt{Line 31}.


\subsection{Adaptive Attacker}
\label{subsec:Adaptive Attacker}

To evaluate the robustness of the classifier against adaptive attacker,
we define a new MILP attack objective, to minimize the following cost:
\vspace{-5pt}
\begin{equation}
\label{eq:gencostobj}
\text{minimize} \sum_{j}a_j w_{x^j} |\tilde{x}^j - x^j| + \sum_{j}(1-a_j) w'_{x^j} |\tilde{x}^j - x^j|
\end{equation}

Where $a_j$ is defined as the following:
\vspace{-5pt}
\begin{equation}
a_j = \begin{cases}
               0               & \tilde{x}^j \leq x^j \\
               1               & \tilde{x}^j > x^j \\
               \end{cases}
\end{equation}

The adaptive attacker wants to minimize the total feature manipulation cost to
generate adversarial example $\tilde{x}$ by perturbing $x$.
We model the total cost as the weighted sum of absolute feature value differences, with different
weights for the increase and decrease changes. Each weight $w_{x^j}$ represents
the unit cost (e.g., some dollar amount) for the attacker to increase feature $x^j$, and $w'_{x^j}$
to decrease it.

To target the box cost constraint, we define $U_N$, $U_L$, $U_M$, and $U_H$ as the sets of feature dimensions
with negligible, low, medium, and high cost to increase, respectively.
We define $V_N $, $V_L$, $V_M$, and $V_H$ as the sets of feature dimensions
with negligible, low, medium, and high cost to decrease, respectively. The adaptive
attacker minimizes the following total feature manipulation cost:
\vspace{-5pt}
\begin{equation}
\label{eq:costobj}
\sum_{k}\sum_{U_k} w_k |\tilde{x}^j - x^j| + \sum_{k}\sum_{V_k} w_k |\tilde{x}^j - x^j|,
k \in \{N, L, M, H\}
\end{equation}

We set weights $w_k$ based on the inverse proportion
of the box for each feature dimension, such that a larger weight prefers a smaller feature change
in the attack. For example, if we allow perturbing a low cost feature to be twice the amount of a
medium cost feature ($\beta = 2 * \gamma$) in the cost-driven constraint, we set $w_L = \frac{1}{2} w_M$, which makes
the adaptive attacker aware that the cost of changing one unit of a medium cost feature is equivalent to changing
two units of a low cost feature in the linear objective.
This adapts the strongest whitebox attack by
including the knowledge of box contraint used in the training.

\section{Evaluation}

In this section, we first evaluate the effectiveness of our core training algorithm
(Section~\ref{sec:Training Algorithm Evaluation}) against the state-of-the-art
robust and regular training methods, and then we evaluate
the end-to-end robust training technique on a security task, Twitter spam detection (Section~\ref{sec:Twitter Spam Detection Application}).

\subsection{Implementation}

We implement our robust training algorithm in XGBoost~\cite{chen2016xgboost} and scikit-learn~\cite{sklearn}.
Our implementation in XGBoost works with all their supported
differentiable loss functions for gradient boosted decision trees
as well as random forest. For scikit-learn, we implement
the robust training algorithm in random forest using the Gini impurity score.

\subsection{Training Algorithm Evaluation}
\label{sec:Training Algorithm Evaluation}

Since the state-of-the-art training method~\cite{chen2019robust} does not support
integrating domain knowledge, we compare our core training algorithm (Algorithm~\ref{alg:robustsplitter}) with $L_\infty$-norm cost model
against existing work without any domain knowledge related cost modeling in this section.
Even though it is unfair to our technique, the experiments in this section act as an ablation study
to show the improvements our Algorithm~\ref{alg:robustsplitter} makes to solve Equation~\eqref{eq:max_uncertain}.
Same as~\cite{chen2019robust}, we run our Algorithm~\ref{alg:robustsplitter} to train
$L_\infty$-norm bounded robustness.


\noindent\textbf{$L_\infty$ robustness definition.}
When the objective of the MILP attack (Section~\ref{sec:Evading Tree Ensembles}) is to minimize the $L_\infty$ distance,
the attack provides the \emph{minimal $L_\infty$-norm evasion distance}
that the attacker needs to perturb in the features in order to evade the model.
In non-security related applications, a larger $L_\infty$ robustness distance
means that a model is more robust. For example, if the MNIST classifier
requires an average of $0.3$ $L_{\infty}$-norm distance changes in adversarial examples, it is more robust than a model with $0.06$
$L_{\infty}$-norm robustness distance, because
the adversarial examples look more differently from the original image.

\noindent\textbf{Accuracy cutoff.} In order to reproduce existing results
in related work~\cite{chen2019robust},
we use $0.5$ prediction confidence as the cutoff
to compute the accuracy scores for all trained models.

\begin{table}[!bt]
	\centering
	\small
	\begin{tabular}{c|r|r|r|r}
		\hline
		\textbf{Dataset} & \begin{tabular}[c]{@{}c@{}}\textbf{Train}\\\textbf{set size}\end{tabular} & \begin{tabular}[c]{@{}c@{}}\textbf{Test}\\\textbf{set size}\end{tabular}  & \begin{tabular}[c]{@{}c@{}}\textbf{Majority}\\\textbf{Class (\%)}\end{tabular} & \begin{tabular}[c]{@{}c@{}}\textbf{\# of}\\\textbf{features}\end{tabular}\\\hline\hline
		breast-cancer & 546 & 137 & 62.64, 74.45 & 10 \\\hline
		cod-rna & 59,535 & 271,617 & 66.67, 66.67 & 8 \\\hline
		ijcnn1 & 49,990 & 91,701 & 90.29, 90.50 & 22 \\\hline
		MNIST 2 vs. 6 & 11,876 & 1,990 & 50.17, 51.86 & 784 \\\hline
	\end{tabular} 
	\caption{Training and testing set sizes, the percentage of majority class in the training and testing set, respectively, and the number of features for the four benchmark datasets.}
	\label{tab:datasets}
\end{table}

\subsubsection{Benchmark Datasets}
We evaluate the robustness improvements in 4 benchmark datasets: breast cancer,
cod-rna, ijcnn1, and binary MNIST (2 vs. 6). Table~\ref{tab:datasets} shows
the size of the training and testing data, the percentage of majority class in
the training and testing set, respectively, and the number of features for these datasets.
We describe the details of each benchmark dataset below.

\begin{table*}[!hbt]
	\centering
	\tabcolsep=1.8pt
	\small

	\begin{tabular}{|c|c|cc|ccc|ccc|ccc|ccc|cc|}
		\hline
		\multirow{2}{*}{Dataset} &\multirow{2}{*}{\begin{tabular}[c]{@{}c@{}}\# of\\trees\end{tabular}} & \multicolumn{2}{c|}{Trained $\epsilon$} & \multicolumn{3}{c|}{Tree Depth} & \multicolumn{3}{c|}{Test ACC (\%)} & \multicolumn{3}{c|}{Test FPR (\%)}  & \multicolumn{3}{c|}{Avg. $l_\infty$} & \multicolumn{2}{c|}{Improv.} \\
		& & Chen's & ours & natural & Chen's & ours & natural & Chen's & ours  & natural & Chen's & ours & natural & Chen's & ours & natural & Chen's \\\hline
		breast-cancer & 4 & 0.30 & 0.30 & 6 & 8 & 8 & 97.81 & 96.35 & 99.27 & 0.98 & 0.98 & 0.98 & .2194 & .3287 & \textbf{.4405} & \textbf{2.01x} & \textbf{1.34x}\\\hline
		cod-rna & 80 & 0.20 & 0.035 & 4 & 5 & 5 & 96.48 & 88.08 & 89.64 & 2.57 & 4.44 & 7.38 & .0343 & .0560 & \textbf{.0664}  & \textbf{1.94x} & \textbf{1.19x}\\\hline
		ijcnn1 & 60 & 0.20 & 0.02 & 8 & 8 & 8 & 97.91 & 96.03 & 93.65 & 1.64 & 2.15 & 1.62 & .0269 & .0327 & \textbf{.0463} & \textbf{1.72x} & \textbf{1.42x}\\\hline
		MNIST 2 vs. 6 & 1,000 & 0.30 & 0.30 & 4 & 6 & 6 & 99.30 & 99.30 & 98.59 & 0.58 & 0.68 & 1.65 & .0609 & .3132 & \textbf{.3317} & \textbf{5.45x} & \textbf{1.06x} \\\hline
	\end{tabular}
	
	\caption{Test accuracy and robustness of GBDT models trained by our algorithm (ours), compared to regularly trained models (natural) and the models trained by Chen and Zhang et al.'s method~\cite{chen2019robust} (Chen's), in XGBoost. The improvement (Improv.) here denotes the average $l_\infty$ robustness distance on our models over regularly trained ones and Chen and Zhang's,
		by measuring adversarial examples found by Kantchelian’s MILP attack~\cite{kantchelian2016evasion}, the strongest whitebox attack.}
	\label{tab:gbdt}
\end{table*}

\noindent\textbf{breast cancer.} The breast cancer dataset~\cite{breast_cancer} contains 2 classes of samples, each representing benign and malignant cells. The attributes represent different measurements of the cell's physical properties (e.g., the uniformity of cell size/shape).

\noindent\textbf{cod-rna.} The cod-rna dataset~\cite{cod-rna} contains 2 classes of samples representing sequenced genomes, categorized by the existence of non-coding RNAs. The attributes contain information on the genomes, including total free-energy change, sequence length, and nucleotide frequencies.

\noindent\textbf{ijcnn1.} The ijcnn1 dataset~\cite{ijcnn} is from the IJCNN 2001 Neural Network Competition. Each sample represents the state of a physical system at a specific point in a time series, and has a label indicating ``normal firing" or ``misfiring". We use the 22-attribute version of ijcnn1, which won the competition. The dataset has highly unbalanced class labels. The majority class in both train and test 
sets are 90\% negatives.

\noindent\textbf{MNIST 2 vs. 6.} The binary mnist dataset~\cite{mnist} contains handwritten digits of ``2" and ``6". The attributes represent the gray levels on each pixel location.

\noindent\textbf{Hyperparameters validation set.} In the hyperparameter
tuning experiments, we randomly separate the training set from the original
data sources into a 90\% train set and a 10\% validation set. We use the
validation set to evaluate the performance of the hyperparamters, and then
train the model again using selected hyperparameters using the entire training set.

\noindent\textbf{Robustness evaluation set.}
In order to reproduce existing results, we follow the same experiment settings used in~\cite{chen2019robust}.
We randomly shuffle the test set, and generate advesarial examples
for 100 test data points for breast cancer, ijcnn1, and binary MNIST, and 5,000 test points
for cod-rna.

\begin{figure}[t!]
	\centering
	\includegraphics[width=0.95\columnwidth]{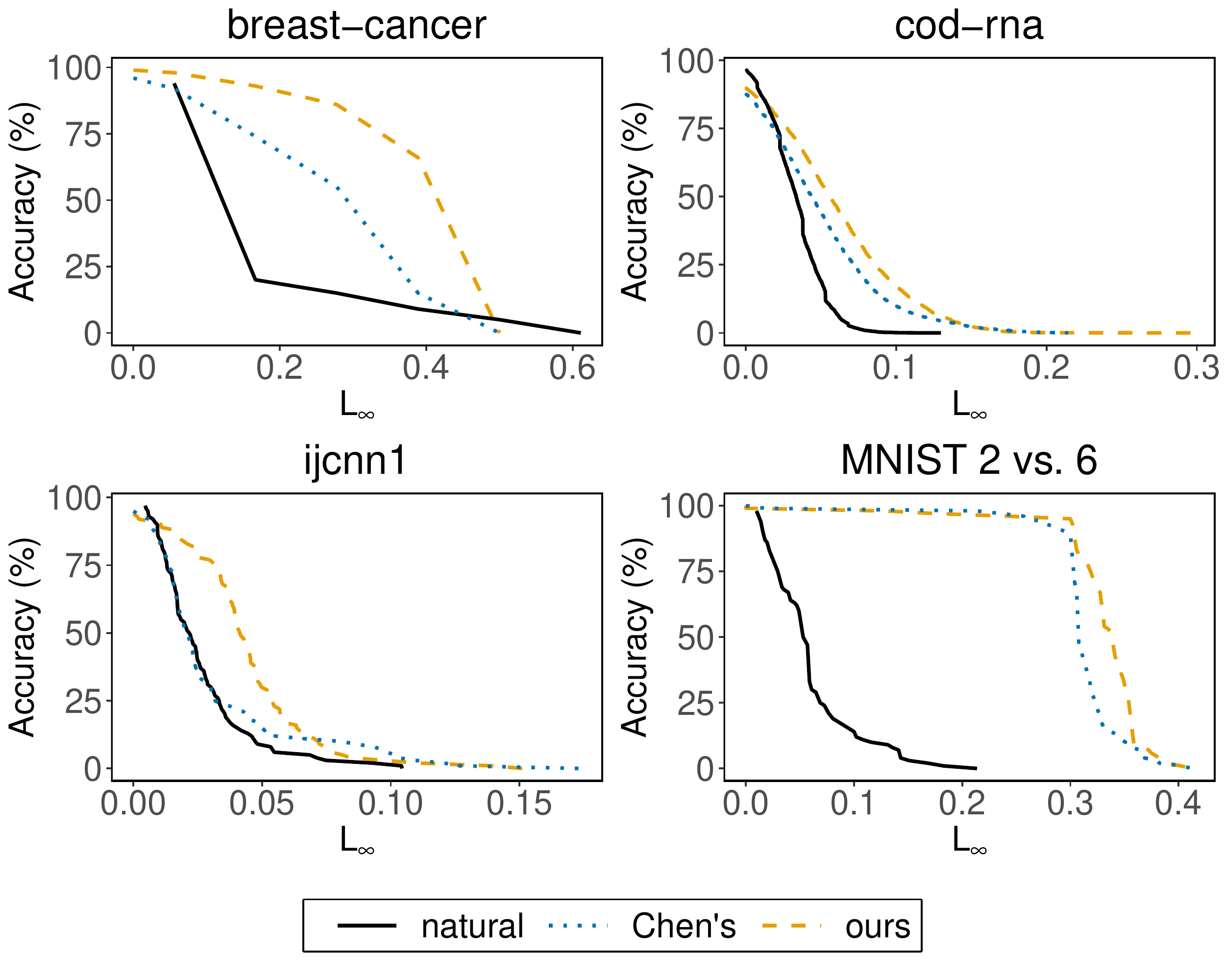}
	\caption{Accuracy under attack.}
	\label{fig:acc_linf_multiplot}
\end{figure}

\subsubsection{GBDT Results}
\label{subsec:gbdt}

We first evaluate the robustness of our training algorithm on the gradient boosted decision trees (GBDT)
using the four benchmark datasets. We measure the model robustness using the $L_\infty$ evasion distance of the adversarial examples found by Kantchelian et al.'s MILP attack~\cite{kantchelian2016evasion}, the strongest whitebox attack that minimizes $L_p$-norm evasion distance for tree ensemble models. We compare the robustness achieved by our algorithm against regular training as well as the state-of-the-art robust training algorithm proposed by Chen et al.~\cite{chen2019robust}. 

\noindent\textbf{Hyperparameters.}
To reproduce the results from existing work~\cite{chen2019robust} and
conduct a fair comparison, we report results from the same number
of trees, maximum depth, and $\epsilon$ for $L_\infty$-norm bound
for regular training and Chen's method in Table~\ref{tab:gbdt}.
For our own training method,
we reused the same number of trees and maximum depth as Chen's method.
Then, we conducted grid search of different $\epsilon$ values by $0.01$
step size to find the model with the best accuracy.
For our cod-rna model, we also experimented with $\epsilon$ values by $0.001$ step size.
To further evaluate their choices of hyperparameters, we have conducted
grid search for the number of trees and maximum depth to measure
the difference in model accuracy, by training 120 models in total.
Our results show that the hyperparameters used in existing work~\cite{chen2019robust} produced similar accuracy as
the best one (Appendix~\ref{appendix:params_gbdt}). Note that the size of the breast-cancer dataset is
very small (only 546 training data points), so using only four trees does not overfit the dataset.

\begin{table*}[!hbt]
	\centering
	\tabcolsep=1.8pt
	\small
	
	\begin{tabular}{|c|cc|ccc|ccc|ccc|ccc|cc|}
		\hline
		\multirow{2}{*}{Dataset} & \multicolumn{2}{c|}{Trained $\epsilon$} & \multicolumn{3}{c|}{Tree Num / Depth} & \multicolumn{3}{c|}{Test ACC (\%)} & \multicolumn{3}{c|}{Test FPR (\%)}  & \multicolumn{3}{c|}{Avg. $l_\infty$} & \multicolumn{2}{c|}{Improv.} \\
		& Chen's & ours & natural & Chen's & ours & natural & Chen's & ours  & natural & Chen's & ours & natural & Chen's & ours & natural & Chen's \\\hline
		breast-cancer & 0.30 & 0.30 & 20 / 4 & 20 / 4 & 80 / 8 & 99.27 & 99.27 & 98.54 & 0.98 & 0.98 & 1.96 & .2379 & .3490 & \textbf{.3872} & \textbf{1.63x} & \textbf{1.11x}\\\hline
		cod-rna & 0.03 & 0.03 & 40 / 14 & 20 / 14 & 40 / 14 & 96.54 & 92.63 & 89.44 & 2.97 & 3.65 & 5.69 & .0325 & .0512 & \textbf{.0675} & \textbf{2.08x} & \textbf{1.32x}\\\hline
		ijcnn1 & 0.03 & 0.03 & 100 / 14 & 100 / 12 & 60 / 8 & 97.92 & 93.86 & 92.26 & 1.50 & 0.78 & 0.08 & .0282 & .0536 & \textbf{.1110} & \textbf{3.94x} & \textbf{2.07x}\\\hline
		MNIST 2 vs. 6 & 0.30 & 0.30 & 20 / 14 & 100 / 12 & 100 / 14 & 99.35 & 99.25 & 99.35 & 0.68 & 0.68 & 0.48 & .0413 & .1897 & \textbf{.2661} & \textbf{6.44x} & \textbf{1.40x}\\\hline
	\end{tabular} 
	
	\caption{Test accuracy and robustness of random forest models trained by our algorithm (ours) compared to regularly trained models (natural), in scikit-learn. The improvement (Improv.) here denotes the average $l_\infty$ robustness distance increase.}
	\label{tab:sklearnrf}
\end{table*}

\noindent\textbf{Minimal evasion distance.} As shown in Table ~\ref{tab:gbdt}, our training algorithm can obtain stronger robustness
than regular training and the state-of-the-art robust training method. On average, the MILP attack needs \gbdtbaselinetimes{} larger $L_\infty$ perturbation distance to evade our models than regularly trained ones. Compared to the state-of-the-art Chen and Zhang et al.'s robust training method~\cite{chen2019robust}, our models require on average \gbdtrobusttimes{} larger $L_\infty$ perturbation distances. Note that the robustness improvement of our trained models are limited on binary MNIST dataset. This is because the trained and tested robustness ranges $L_\infty\leq 0.3$ are fairly large for MNIST dataset. The adversarial examples beyond that range are not imperceptible any more, and thus the robustness becomes extremely hard to achieve without heavily sacrificing regular accuracy.

\begin{figure}[t!]
	\centering
	\includegraphics[width=0.48\columnwidth]{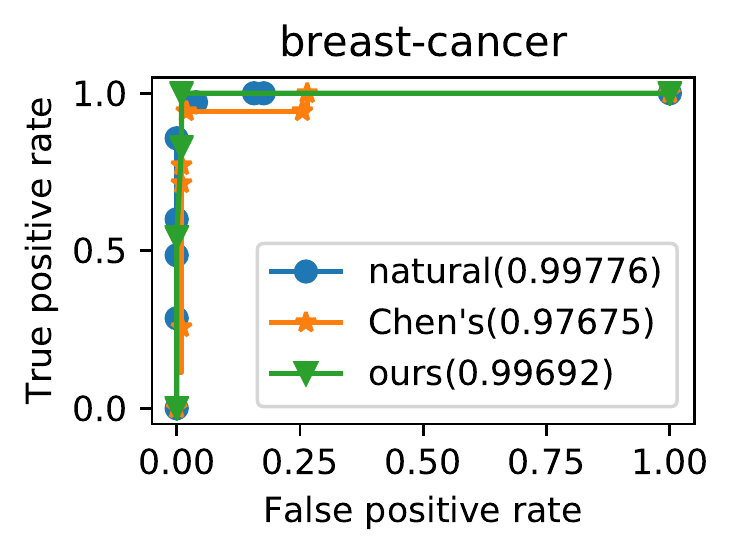}
	\includegraphics[width=0.48\columnwidth]{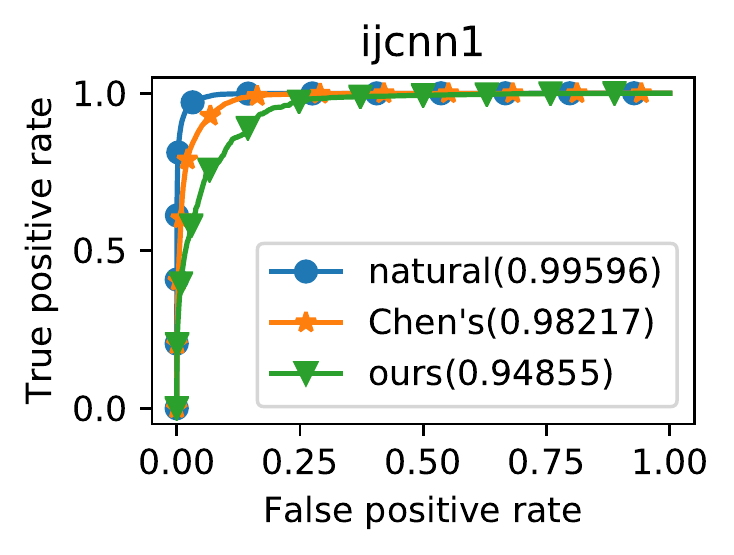}
	\includegraphics[width=0.48\columnwidth]{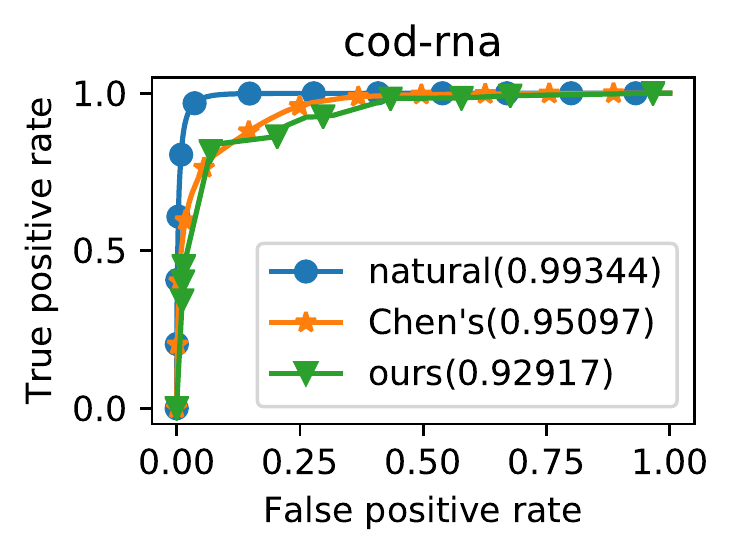}
	\includegraphics[width=0.48\columnwidth]{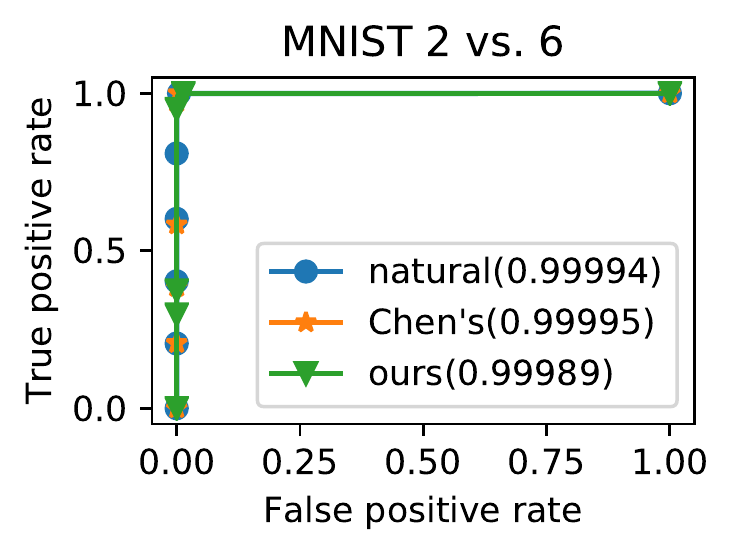}
	\caption{ROC curves of GBDT models trained with natural, Chen's, and our training methods. AUC is given in the legend.}
	\label{fig:gbdt-aoc}
\end{figure}

\noindent\textbf{Accuracy under attack.}
Using the minimal $l_\infty$ evasion distances of adversarial examples,
we plot how the accuracy of the models decrease as the attack distance
increases in Figure~\ref{fig:acc_linf_multiplot}.
Compared to regular training, our models maintain higher accuracy
under attack for all datasets except the breast-cancer one.
Both breast-cancer models trained by our method and Chen's method
reach 0\% accuracy at $l_\infty$ distance 0.5, whereas
the largest evasion distance
for the regularly trained model is 0.61.
For the breast-cancer dataset, robustly trained models have an larger evasion distance than the regularly trained model for over 94\% data points.
Compared to Chen's method, our models maintain higher accuracy under attack in
all cases, except a small distance range for the
ijcnn1 dataset ($l_\infty$ from 0.07 to 0.10, Figure~\ref{fig:acc_linf_multiplot}).

\noindent\textbf{Model quality evaluation.}
\label{subsec:roc}
Figure~\ref{fig:gbdt-aoc} shows the ROC curves and the Area Under the Curve (AUC) for GBDT models trained with natural, Chen's, and our training methods. For all the four testing datasets, the AUC of the model trained by our method is on par with the other two algorithms. On average, AUC of the model trained by our method is only 0.03 and 0.01 lower, while our method can increase the MILP attack cost by \gbdtbaselinetimes{} and \gbdtrobusttimes{} than natural and Chen's training methods, respectively.
Table~\ref{tab:gbdt} shows that overall we maintain
relatively high accuracy and low false positive rate.
However, we have a high false positive rate for the cod-rna model and low accuray for the ijcnn1 model,
as the tradeoff to obtain stronger robustness.

\subsubsection{Random Forest Results}
\label{subsec:rf}

We evaluate the robustness of random forest models trained with our algorithm
on the four benchmark datasets. We compare against Chen's algorithm~\cite{chen2019robust} and
regular training in scikit-learn~\cite{sklearn}.
Since Chen's algorithm is not available in scikit-learn, we have implemented their algorithm to train random forest models ourselves.
We compare the effectiveness of our robust training algorithm
against Chen's method and regular training, when using
Gini impurity to train random forest scikit-learn.

\noindent\textbf{Hyperparameters.}
We conduct a grid search for the number of trees and maximum depth hyperparameters.
Specifically, we use the following number of trees:
20, 40, 60, 80, 100, and the maximum depth: 4, 6, 8, 10, 12, 14.
For each dataset, we train 30 models, and select the hyperparameters
with the highest validation accuracy. The resulting hyperparameters
are shown in Table~\ref{tab:sklearnrf}.
For the breast-cancer and binary mnist datasets,
we re-used the same $\epsilon = 0.3$ from Chen's
GBDT models.
We tried different $L_\infty \leq \epsilon$ values of robust training for cod-rna and ijcnn1 datasets,
and found out that $\epsilon = 0.03$ gives a reasonable
tradeoff between robustness and accuracy. For example,
when $\epsilon = 0.2$, we trained 30 random forest models using Chen's method for the cod-rna dataset, and the best validation accuracy is only $79.5\%$. Whereas, using $\epsilon=0.03$ increases
the validation accuracy to $91\%$.

\begin{figure}[t!]
	\centering
	\includegraphics[width=0.95\columnwidth]{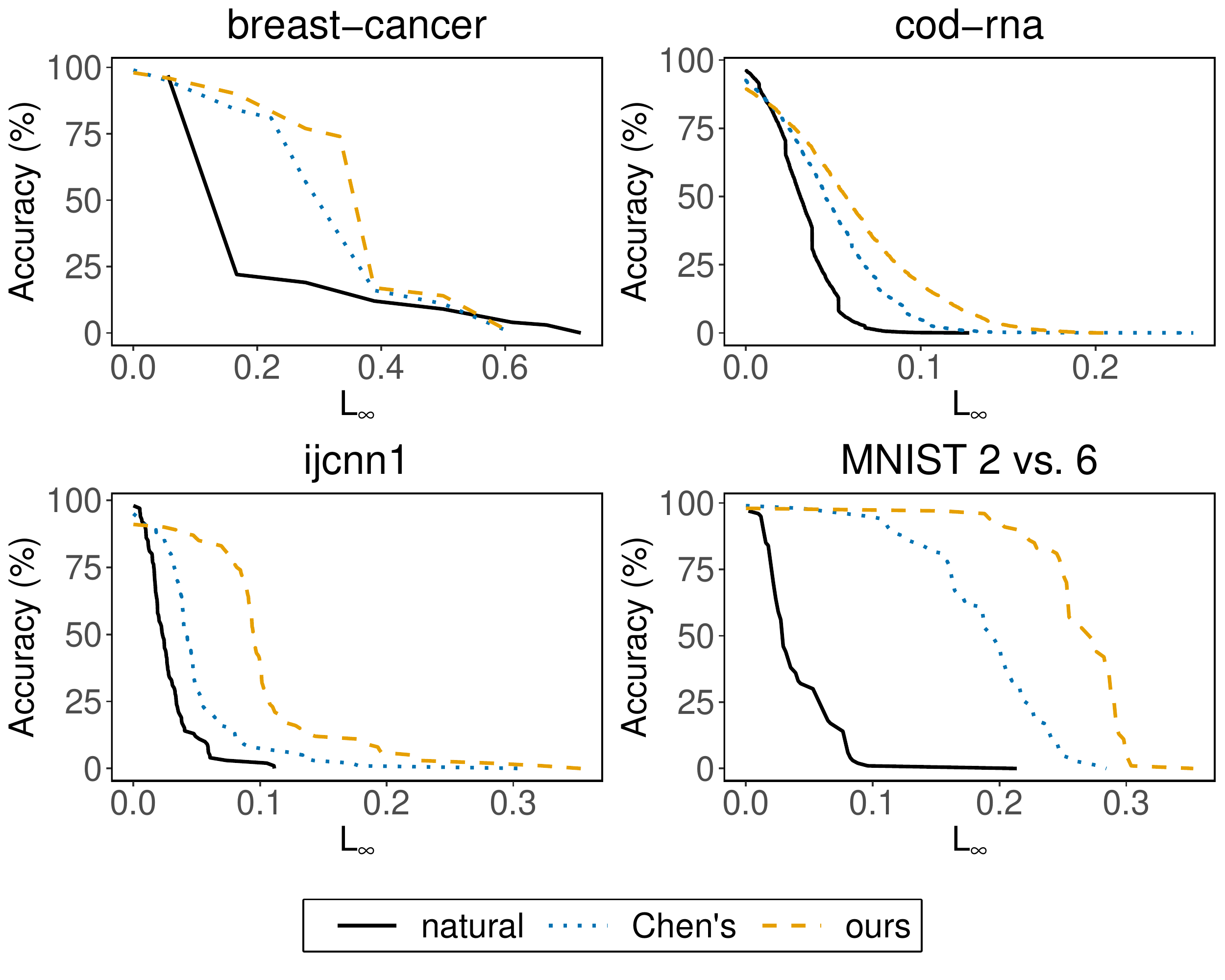}
	\caption{Accuracy under attack.}
	\label{fig:acc_linf_sklearnrf_multiplot}
\end{figure}

\begin{figure}[t!]
	\centering
	\includegraphics[width=0.48\columnwidth]{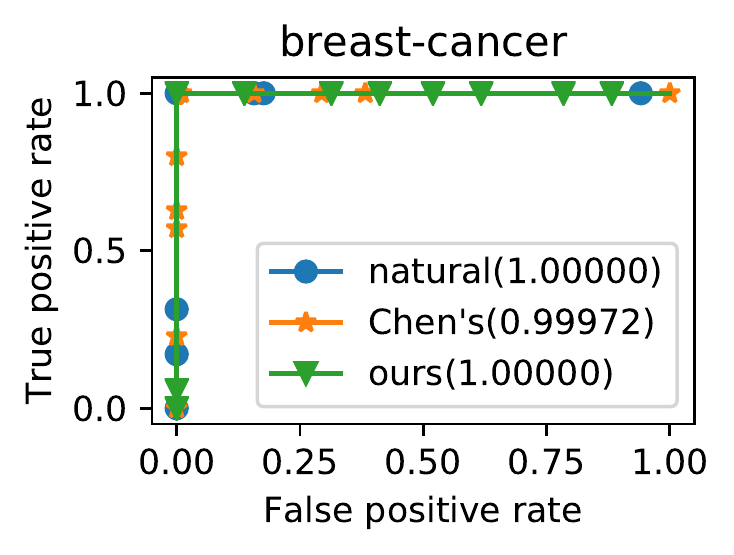}
	\includegraphics[width=0.48\columnwidth]{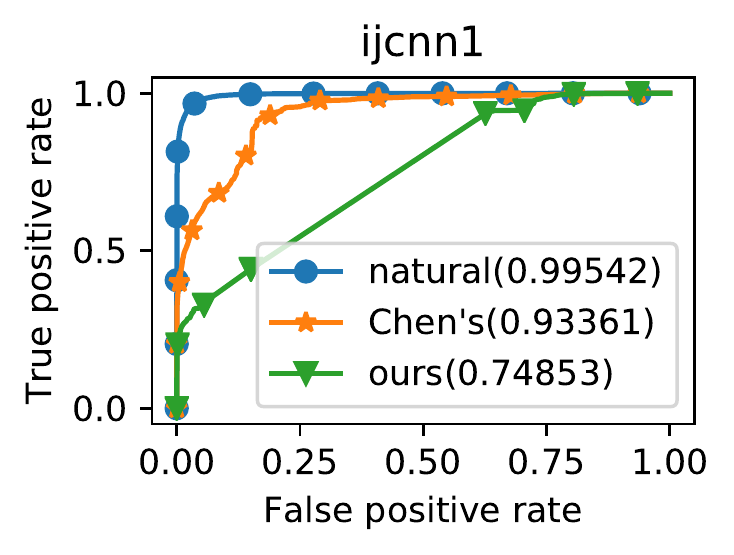}
	\includegraphics[width=0.48\columnwidth]{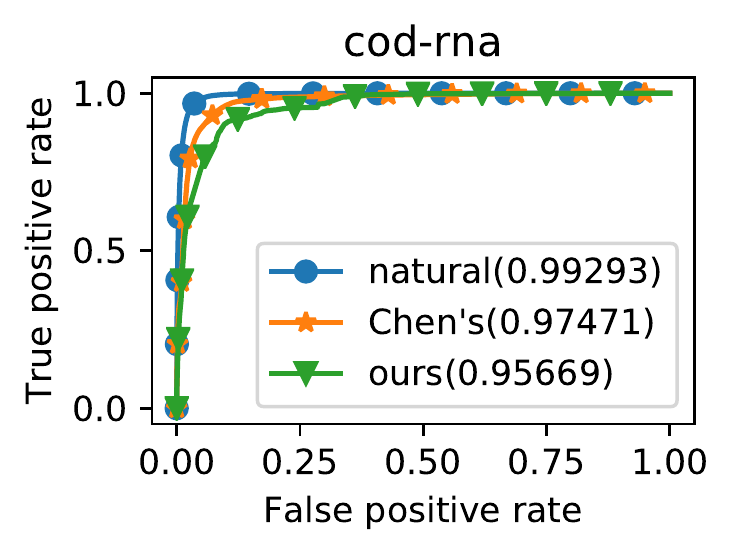}
	\includegraphics[width=0.48\columnwidth]{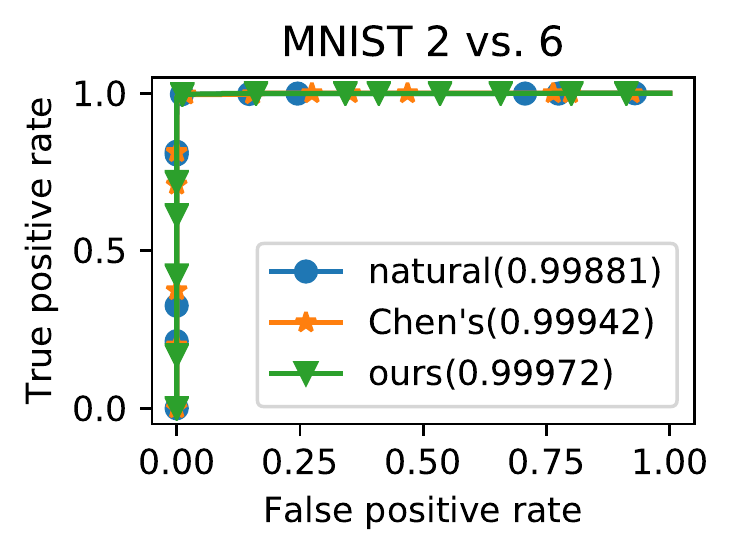}
	\caption{ROC curves of random forest models trained with natural, Chen's, and our training methods in scikit-learn. AUC is given in the legend.}
	\label{fig:sk-rf-aoc}
\end{figure}

\noindent\textbf{Minimal Evasion Distance.}
As shown in Table~\ref{tab:sklearnrf}, the robustness of our random forest models outperforms the ones from regular training and Chen's algorithm. Specifically, the average $l_\infty$ distance of adversarial examples against our robust models found by Kantchelian et al.'s MILP attack~\cite{kantchelian2016evasion} is on average \sklearnrfbaselinetimes{} and \sklearnrfrobusttimes{} larger than regular training and Chen's method. On the other hand, there is only a $1.35\%$ average drop of test accuracy and under $3\%$ increase of false positive rate for the robust models compared to Chen's method.

\noindent\textbf{Accuracy under attack.}
Figure~\ref{fig:acc_linf_sklearnrf_multiplot} shows the accuracy of models under different $L_\infty$ evasion distances
of the MILP attack. Our models maintain higher accuracy under attack than those trained by Chen's method
for all datasets. In addition, our models maintain higher accuracy under attack than regular training,
for all datasets except a small region for the breast-cancer dataset ($L_\infty > 0.5$).

\noindent\textbf{Model quality evaluation.}
Figure~\ref{fig:sk-rf-aoc} shows the ROC curves and the Area Under the Curve (AUC) for
random forest models trained with natural, Chen's, and our training methods.
For three datasets (breast-cancer, cod-rna, and binary MNIST), the ROC curve of our
models are very close to the baseline models, with at most 0.018 drop in AUC than
Chen's method. However, our random forest model for the ijcnn1 dataset has very low AUC (0.74853).
The model has 92\% test accuracy, and the majority class of
the test set is 90\% negative class. Note that the false positive rate for the model is only
0.08\% because the model does not predict the positive class very often, and therefore it
generates very few false positives.
We acknowledge the limitation of applying our algorithm in the $L_\infty$ norm cost
model for the ijcnn1 dataset. This also motivates the need for cost models other than
the $L_\infty$ norm. In Section~\ref{sec:results}, we demonstrate that we can balance
robustness and accuracy using a cost model consistent with the semantics of the features
for Twitter spam detection, even though using the $L_\infty$ cost model significantly degrades
the model quality.

\subsubsection{Benefits of our robust algorithm over existing heuristics}
\label{section:heuristic_benefit}


According to Equation~\eqref{eq:max_uncertain}, our robust algorithm is designed to maximize some impurity measure for each potential feature split during the training process. The higher the score is obtained by the algorithm, the stronger capability of the attacker is used for training, which guides the model to learn stronger robustness. Therefore, how well the algorithm can solve the maximization problem directly determines the eventual robustness the models can learn. To that end, we measure how our robust algorithm performs in solving the maximization problem compared to the heuristics used in state-of-the-art Chen and Zhang's~\cite{chen2019robust} to illustrate its effectiveness. 

\begin{table}[!hbt]
	\centering
	\tabcolsep=3.5pt
	\small
	\begin{tabular}{c|r|r|r|r}
		\hline
		\textbf{Dataset} & \textbf{Better (\%)} & \textbf{Equal (\%)}   & \textbf{Worse} (\%) & \textbf{Total}\\\hline\hline
		breast-cancer & 99.74 & 0.26 & 0 & 3,047 \\\hline
		cod-rna & 94.13 & 4.66 & 1.21 & 35,597 \\\hline
		ijcnn1 & 90.31 & 1.11 & 8.58 & 424,825\\\hline
		MNIST 2 vs. 6 & 87.98 & 6.33 & 5.69 & 796,264\\\hline
	\end{tabular} 
	\caption{The percentage of the cases where our robust algorithm performs better, equally well, or worse than the heuristics used in the state-of-the-art Chen and Zhang et al.'s robust training algorithms~\cite{chen2019robust} in solving the maximization problem (Equation~\ref{eq:max_uncertain}). The total number of cases represent the total number of splits evaluated during robust optimization.}
	\label{tab:count}
\end{table}

On the four benchmark datasets, we measure the percentage of the cases where our robust algorithms can better solve the maximization problem than the heuristics used in~\cite{chen2019robust} and summarize the results in Table~\ref{tab:count}. The results show that our robust algorithm can provide a better solution than heuristics used in Chen and Zhang et al.'s method~\cite{chen2019robust} for at least 87.98\% cases during the whole training process. On small datasets like breast-cancer and cod-rna, our algorithm performs equally or better for 100\% and 98.79\% cases respectively. Such significant improvements in solving the maximizaation problem greatly benefit the robustness of our trained models. The results provide insights on why our robust training algorithm can obtain more robust tree ensembles than existing training methods.

\subsection{Twitter Spam Detection Application}
\label{sec:Twitter Spam Detection Application}

In this section, we apply our robust tree ensemble training method to a classic security application,
spam URL detection on Twitter~\cite{kwon2017domain}. As a case study,
we want to answer the following questions in the evaluation:

\begin{itemize}
\setlength\itemsep{0em}
\item \textbf{Cost-driven constraint:} How to specify the cost-driven constraint based on security domain knowledge?
What is the advantage of training cost-driven constraint compared to $L_\infty$-norm robustness?
\item \textbf{Accuracy vs robustness tradeoffs:} How much
does robustness affect accuracy? Do different ways
of specifying the cost-driven constraint change that tradeoffs?
\item \textbf{Adaptive attack cost:} Against the strongest whitebox attack~\cite{kantchelian2016evasion},
does the robust model increase the adaptive attack cost for successful evasion?
\item \textbf{Other mathematical distances:} Can we increase robustness against
$L_1$ and $L_2$ based attacks?
\end{itemize}


\subsubsection{Dataset}

We obtain the public dataset used in Kwon et al.'s work~\cite{kwon2017domain}
to detect spam URLs posted on Twitter. Spammers spread harmful URLs
on social networks such as Twitter to distribute malware, scam, or phishing content.
These URLs go through a series of redirections, and eventually reach a landing
page containing harmful content. The existing detectors proposed in prior works often make decisions based on content-based features
that are strong in predictive power but easy to be changed, e.g., different words
used in the spam tweet. Kwon et al. propose to use more robust features that
incur monetary or management cost to be changed under adversarial settings.
They extract these robust features from the URL redirection chains (RC)
and the corresponding connected components (CC) formed by the chains.

\begin{table}[ht!]
  \centering
  \small
  \begin{tabular}{l | r | r}
    \hline
    \textbf{Dataset} & \textbf{Training} & \textbf{Testing} \\
       \hline
       \hline
       Malicious & 130,794 & 55,732 \\
       Benign & 165,076 & 71,070\\
       \hline
       Total & 295,870 & 126,802 \\
       \hline
  \end{tabular}
  \caption{
  	The size of Twitter spam dataset~\cite{kwon2017domain}.} 
  \label{tab:twitterdata}
\end{table}


\begin{table}[ht!]
  \centering
  \footnotesize
  \begin{tabular}{l | l | l | l}
    \hline
    \textbf{Feature Name} & \textbf{Description} & \multicolumn{2}{|c}{\textbf{Cost}} \\
    & & \multicolumn{1}{l}{$\downarrow$} & $\uparrow$  \\
       \hline
       \hline
       \multicolumn{4}{l}{Shared Resources-driven Features} \\
       \hline
       EntryURLid & In degree of the largest redirector & M & N  \\
       AvgURLid & \begin{tabular}[l]{@{}l@{}}{Average in degree of}\\{URL nodes in the RC}\end{tabular} & M & N  \\
       ChainWeight & Total frequency of edges in the RC & L & N  \\
       CCsize &  \# of nodes in the CC & L & N  \\
       CCdensity & Edge density of the CC & L & N  \\
       MinRCLen & Min length of the RCs in the CC & L & N  \\
       AvgLdURLDom & \begin{tabular}[l]{@{}l@{}}{Average domain \# of}\\{landing URL IPs in the CC}\end{tabular} & \textbf{H} & N \\
       AvgURLDom & \begin{tabular}[l]{@{}l@{}}{Average domain \# for}\\{the IPs in the RC}\end{tabular} & M & N  \\
       \hline
       \multicolumn{4}{l}{Heterogeneity-driven Features} \\
       \hline
       GeoDist & \begin{tabular}[l]{@{}l@{}}{Total geo distance (km)}\\{traversed by the RC}\end{tabular} & \textbf{H} & N  \\
       CntContinent &  \# of unique continents in the RC & M & N  \\
       CntCountry &  \# of unique countries in the RC & M & N  \\
       CntIP &  \# of unique IPs in the RC & L & N  \\
       CntDomain & \# of unique domains in the RC & L & N  \\
       CntTLD & \# of unique TLDs in the RC & L & N  \\
       \hline
       \multicolumn{4}{l}{Flexibility Features} \\
       \hline
       ChainLen & Length of the RC & L & N  \\
       EntryURLDist & \begin{tabular}[l]{@{}l@{}}{Distance from the initial URL}\\{to the largest redirector}\end{tabular} & L & N  \\
       CntInitURL & \# of initial URLs in the CC & L & N  \\
       CntInitURLDom & \begin{tabular}[l]{@{}l@{}}{Total domain name \#}\\{in the initial URLs}\end{tabular} & L & N  \\
       CntLdURL & \# of final landing URLs in the RC & L & N  \\
       AvgIPperURL & Average IP \# per URL in the RC & L & N  \\
       AvgIPperLdURL & \begin{tabular}[l]{@{}l@{}}{Average IP \# per}\\{landing URL in the CC}\end{tabular} & L & \textbf{H}  \\
       \hline
       \multicolumn{4}{l}{User Account Features} \\
       \hline
       Mention Count & \# of `@' count to mention other users & N & L  \\
       Hashtag Count & \# of hashtags & N & L  \\
       Tweet Count & \# of tweets made by the user account & N & M  \\
       URL Percent & \begin{tabular}[l]{@{}l@{}}{Percentage of user posts}\\{that contain a URL}\end{tabular} & N & L  \\
       \hline
    \multicolumn{4}{l}{$*$ CC: connected component. RC: redirection chain.}\\
    \multicolumn{4}{l}{BPH: bulletproof hosting. N: Negligible. L: Low. M: Medium. H: High.}
  \end{tabular}
  \caption{
  	We reimplement 25 features used in~\cite{kwon2017domain} to detect Twitter spam, among which three features have high cost to decrease or increase. To maintain the same level of spam activities, the attacker needs to purchase more bulletproof hosting servers to host the different landing pages if AvgLdURLDom feature is decreased or AvgIPperLdURL feature is increased. In addition, it is very hard for the attacker to decrease the GeoDist feature.
	}
  \label{tab:twitter_spam_features}
\end{table}

\textbf{Feature extraction.} We reimplemented and extracted 25 features from the dataset in the original paper,
as shown in Table~\ref{tab:twitter_spam_features}. There are four families of features:
shared resources-driven, heterogeneity-driven, flexibility-driven, and user account and post level features.
The key intuitions behind the features are as follows. 1) Attackers reuse underlying hosting infrastructure to reduce the economic cost of renting and maintaining servers. 2) Attackers use machines hosted on bulletproof hosting services or compromised machines to operate the spam campaigns.
These machines are located around the world, which tend to spread
over larger geographical distances than benign hosting infrastructure, and it is hard for attackers to control
the geographic location distribution of their infrastructure. 3) Attackers want to maximize the flexibility of
the spam campaign, so they use many different initial URLs to make the posts look distinct, and
different domains in the long redirection chains to be agile against takedowns. 4) Twitter spammers
utilize specific characters to spread harmful content, such as hashtags and `@' mentions.
We removed some highly correlated features from the original paper. For example,
for a feature where the authors use both maximum and average numbers, we use the average
number only.

Kwon et al. labeled the dataset by crawling suspended users, identifying benign users,
and manually annotating tweets and URLs. In total, there are 186,526 distinct malicious tweets
with spam URLs, and 236,146 benign ones. We randomly split the labeled dataset into
70\% training set and 30\% testing set as shown in Table~\ref{tab:twitterdata}.
We extract the aforementioned 25 features from each data point and normalize the values to be between 0 and 1
for training and testing.



\subsubsection{Attack Cost Analysis}

In order to obtain the cost-driven constraint for robust training,
we first analyze the cost of changing the features and the direction of
the changes, then we specify a box contraint for the cost accordingly.

\paragraph{Feature Analysis}

We categorize the features into negligible (N), low (L), medium (M), and high (H) cost to change, as shown in Table~\ref{tab:twitter_spam_features}.
We analyze the cost based on feature families as follows.

\begin{itemize}[leftmargin=*]
\setlength\itemsep{0em}
\item \textbf{Shared resources:} All features cost more to be decreased than to be increased.
If the attacker decreases the reused redirectors in the chain, the attacker needs to set up
additional redirector servers to maintain the same level of spam activities
(EntryURLid and AvgURLid features). It costs even more to set up more servers for the landing
pages, since the landing URLs contain actual malicious content, which are usually hosted
on bulletproof hosting (BPH) services. Feature AvgLdURLDom captures how the attacker is reusing the
malicious content hosting infrastructure. If the value is decreased, the attacker will need to set up more
BPH severs, which has the highest cost in the category.
\item \textbf{Heterogeneity:} The total geographical distance traversed by the URL nodes in the redirection
chain has the highest cost to change in general (GeoDist). If the attacker uses all the available
machines as resources for malicious activities, it is hard to control the location of
the machines and the distance between them. Overall, it is harder to decrease GeoDist to what looks more
like benign value than to increase it. Since GeoDist values for benign URL redirection chains are
very concentrated in one or two countries, the attacker would need to purchase more expensive resources
located close by to mimic benign URL. The other four features that count number of continents, countries,
IPs, domains, and top-level domains incur cost for decreased flexibility and increased maintainence cost
if the features are decreased.
\item \textbf{Flexibility:} All features in this family except the last one have relatively low cost to decrease,
because that decreases the flexibility of the attack. The high cost feature AvgIPperLdURL counts the number of
IP addresses that host the malicious landing page URL. If the attacker wants more flexibility of hosting
the landing page on more BPH servers, the cost will be increased significantly.
\item \textbf{User account:} Increasing features in this family generally increases suspiciousness
of the user account. Among them, increasing the tweet count is the most suspicious of all,
since a tweet is capped by 140 characters which limits the number of mentions and hashtags,
and percentage of posts containing URLs is also capped. If a user account sends too many tweets
that puts the account to the top suspicious percentile, it can be easily detected by simple filtering
mechanism and compromise the account.
\end{itemize}

Overall, three features have the highest cost to be perturbed: AvgLdURLDom, GeoDist\footnote{GeoDist, CntContinent and CntCountry have similar intuition,
but we choose GeoDist since it has finer granularity in feature values.},
and AvgIPperLdURL. 
Decreasing AvgLdURLDom and increasing AvgIPperLdURL
incurs cost to obtain more bulletproof hosting servers for the landing page URL, and
manipulating GeoDist is generally outside the control of the attacker.
Other types of actions can also achieve the changes in AvgLdURLDom and AvgIPperLdURL,
but it will generally decrease the profit of the malicious operation.
To decrease AvgLdURLDom, if the attacker does not rent more BPH servers but only reduces
the number of malicious landing pages, that reduces the profit.
If the attacker increases the AvgIPperLdURL by using cheap servers, their malicious content
could be taken down more often that interrupts the malicious operation.

\begin{table*}[ht!]
	\centering
	\small
	\tabcolsep=5.8pt
	\begin{tabular}{c | cccc | c | crc | cccccc}
		\hline
		\multirow{3}{*}{\begin{tabular}[l]{@{}l@{}}{\textbf{Classifier}}\\{\textbf{Model}}\end{tabular}} & \multicolumn{4}{c|}{\textbf{Constraint Variables}} & \multirow{3}{*}{\begin{tabular}[l]{@{}l@{}}{\textbf{Adaptive}}\\{\textbf{Objective}}\end{tabular}} & \multicolumn{3}{c|}{\multirow{2}{*}{\textbf{Model Quality}}} & \multicolumn{6}{c}{\textbf{Robustness against MILP}} \\
		& N & L & M & H &  & & & & \multicolumn{6}{c}{Average} \\
		& \textbf{$\alpha$} & \textbf{$\beta$} & \textbf{$\gamma$} & \textbf{$\mu$} & & Acc & FPR & AUC & $L_1$ & $L_2$ & $Cost_1$ & $Cost_2$ & $Cost_3$ & $Cost_4$ \\
		\hline
		\hline
		Natural & - & - & - & - & - & 99.38 & 0.89 &  .9994 & .007 & .006 & .010 & .009 & .009 & .008 \\ \hline
		C1 $\epsilon=0.03$ & - & - & - & - & - & 96.59 & 5.49 & .9943 & .046 & .036 & .080 & .070 & .062 & .054 \\ \hline
		C2 $\epsilon=0.05$ & - & - & - & - & - & 94.51 & 7.27 & .9910 & .062 & .053 & .133 & .109 & .089 & .085 \\ \hline
		C3 $\epsilon=0.1$ & - & - & - & - & - & 91.89 & 11.96 & .9810 & \textbf{.079} & \textbf{.062} & \textbf{.156} & \textbf{.133} & \textbf{.111} & \textbf{.099} \\ \hline\hline
		M1 & 0.08 & 0.04 & 0.02 & 0 & \multirow{5}{*}{$Cost_1$} & 98.24 & 2.05 & .9984 & .032 & .027 &  .099 & .051 & .058 & .056  \\ \cline{1-5}\cline{7-15}
		M2 & 0.12 & 0.06 & 0.03 & 0 &  & 96.54 & 4.09 & .9941 & \textbf{.043} & \textbf{.036} & \textbf{.106} & \textbf{.078} & \textbf{.064} & \textbf{.062} \\ \cline{1-5}\cline{7-15}
		M3 & 0.20 & 0.10 & 0.05 & 0 &  & 96.96 & 4.10 & .9949 & .033 & .027 & .064 & .025 & .040 & .040 \\ \cline{1-5}\cline{7-15}
		M4 & 0.28 & 0.14 & 0.07 & 0 &  & 94.38 & 4.25 & .9884 & .024 & .012 & .043 & .026 & .039 & .023 \\ \cline{1-5}\cline{7-15}
		M5 & 0.32 & 0.16 & 0.08 & 0 &  & 93.85 & 9.62 & .9877 & .024 & .015 & .034 & .025 & .030 & .025 \\ \hline\hline
		M6 & 0.09 & 0.06 & 0.03 & 0.03 & \multirow{4}{*}{$Cost_2$} & 97.82 & 2.65 & .9968 & \textbf{.049} & .038 & \textbf{.104} & \textbf{.090} & \textbf{.070} & \textbf{.068}  \\ \cline{1-5}\cline{7-15}
		M7 & 0.15 & 0.10 & 0.05 & 0.05 &  & 96.60 & 4.91 & .9929 & .045 & \textbf{.039} & .080 & .072 & .061 & .060  \\ \cline{1-5}\cline{7-15}
		M8 & 0.24 & 0.16 & 0.08 & 0.08 &  & 93.10 & 9.16 & .9848 & .041 & .030 & .082 & .057 & .050 & .049 \\ \cline{1-5}\cline{7-15}
		M9 & 0.30 & 0.20 & 0.10 & 0.10 & & 92.28 & 12.16 & .9836 & .042 & .028 & .050 & .044 & .041 & .038 \\ \hline\hline
		M10 & 0.04 & 0.04 & 0.02 & 0 & \multirow{6}{*}{$Cost_3$}  & 98.51 & 1.84 & .9988 & .025 & .022 & .087 & .041 & .052 & .049 \\ \cline{1-5}\cline{7-15}
		M11 & 0.06 & 0.06 & 0.03 & 0 &  & 97.31 & 3.65 & .9953 & .029 & .017 & .032 & .027 & .026 & .025 \\ \cline{1-5}\cline{7-15}
		M12 & 0.10 & 0.10 & 0.05 & 0 &  & 96.86 & 4.07 & .9919 & .044 & .035 & .062 & .059 & .049 & .048  \\ \cline{1-5}\cline{7-15}
		M13 & 0.16 & 0.16 & 0.08 & 0 &  & 94.54 & 5.91 & .9900 & \textbf{.051} & \textbf{.041} & \textbf{.109} & \textbf{.090} & \textbf{.075} & \textbf{.074}  \\ \cline{1-5}\cline{7-15}
		M14 & 0.20 & 0.20 & 0.10 & 0 &  & 96.36 & 4.95 & .9910 & .033 & .024 & .054 & .042 & .043 & .043 \\ \cline{1-5}\cline{7-15}
		M15 & 0.28 & 0.28 & 0.14 & 0 &  & 93.81 & 6.57 & .9851 & .039 & .039 & .093 & .070 & .048 & .048 \\ \hline\hline
		M16 & 0.06 & 0.03 & 0.03 & 0 &  \multirow{4}{*}{$Cost_4$} & 97.31 & 3.48 & .9953 & .036 & .018 & .038 & .035 & .034 & .028 \\ \cline{1-5}\cline{7-15}
		M17 & 0.10 & 0.05 & 0.05 & 0 &  & 97.41 & 2.70 & .9964 & .023 & .020 & \textbf{.084} & .034 & .051 & .049  \\ \cline{1-5}\cline{7-15}
		M18 & 0.16 & 0.08 & 0.08 & 0 &  & 93.41 & 9.08 & .9872 & .035 & .024 & .074 & \textbf{.044} & \textbf{.062} & \textbf{.051} \\ \cline{1-5}\cline{7-15}
		M19 & 0.20 & 0.10 & 0.10 & 0 &  & 96.48 & 4.75 & .9918 & \textbf{.047} & \textbf{.038} & .054 & .041 & .051 & \textbf{.051} \\ \hline
	\end{tabular}
	\caption[Caption without FN]{
		We trained classifiers with 19 different cost models under the box constraint, and we compare them against regular training (Natural) and three models
		from Chen's method~\cite{chen2019robust} with different $\epsilon$.
		We separate our models by four different cost families. Each cost family
		keeps the same proportion between the constraint variables and has the same adaptive attack objective.
		The best numbers within each cost family are highlighted in bold.
		We have also evaluated the recall of the models in Appendix~\ref{appendix:recall}.}
	\label{tab:costmodels}
\end{table*}

\begin{table}[ht!]
	\centering
	\small
	\begin{tabular}{c | c c c c}
		\hline
		\multirow{2}{*}{\textbf{Objective}} & \multicolumn{4}{c}{\textbf{Adaptive Attack Weights}}  \\
		& \textbf{$w_N$} & \textbf{$w_L$} & \textbf{$w_M$} & \textbf{$w_H$} \\
		\hline
		\hline
		$Cost_1$ & 1 & 2 & 4 & $\infty$ \\ \hline
		$Cost_2$ & 1 & 2 & 3 & 3 \\ \hline
		$Cost_3$ & 1 & 1 & 2 & $\infty$ \\ \hline
		$Cost_4$ & 1 & 2 & 2 & $\infty$ \\ \hline
		\hline
	\end{tabular}
	\caption{
		The weights in adaptive attack objective to target the four different families of cost models
		in Table~\ref{tab:costmodels}.} 
	\label{tab:adapobj}
\end{table}

\subsubsection{Box Constraint Specification}
\label{subsubsect:Box Constraint Specification}

We specify box constraint according to Section~\ref{sec:Box Cost constraint}
with 19 different cost models as shown in Table~\ref{tab:costmodels},
from M1 to M19.
We want to allow more perturbations for lower cost features than higher cost ones,
and more perturbations on the lower cost side (increase or decrease)
than the higher cost side. According to Table~\ref{tab:cost_table} and our analysis
of the features in Table~\ref{tab:twitter_spam_features}, we assign the constraint variables
$\alpha$, $\beta$, $\gamma$ and $\mu$ to negligible, low, medium, and high cost.

We specify four families of cost models, where each one has a corresponding adaptive attack cost
to target the trained classifiers. We assign four distinct constraint variables in the first family (M1 to M5). 
For all the other families (M6 to M9, M10 to M15, and M16 to M19), we assign only
three distinct constraint variables, representing three categories of feature
manipulation cost, by repeating the same value for two out of four categories.
For example, the second cost family (M6 to M9) merges medium and high cost categories
into one using the same value for $\gamma$ and $\mu$.
Within the same cost family, the relative scale of the constraint variables
between the categories are the same, and the adaptive attack cost for that group
is the inverse proportion of the constraint variables, as we have discussed in Section~\ref{subsec:Adaptive Attacker}.
For example, in the second group (M6 to M9), values for the low cost perturbation range ($\beta$)
are twice the amount of the medium cost ones ($\gamma$), and the values for the negligible cost ($\alpha$)
are three times of the medium cost ones ($\gamma$). Therefore, in the adaptive attack objective, we assign
$w_N = 1, w_L = 2, w_M = 3$ to capture that the cost of perturbing one, two, and three units of
negligible, low, and medium cost features are equivalent. For each cost family,
we vary the size of the bounding box to represent different attacker budget during training, resulting in 19 total settings of the constraint.

Using the cost-driven constraint, we train robust gradient boosted decision trees. We compare our training algorithm against regular training and
Chen's method~\cite{chen2019robust}.
We use 30 trees, maximum depth 8, to train one model using regular training.
For Chen's training algorithm,
we specify three different $L_\infty$ norm cost models ($\epsilon =$ 0.03, 0.05, and 0.1)
to obtain three models C1, C2, and C3 in Table~\ref{tab:costmodels}.
For Chen's method and our own algorithm,  we use 150 trees, maximum depth 24.


\subsubsection{Results}
\label{sec:results}

\noindent\textbf{Adaptive attack cost:} Compared to regular training, our best model increases the cost-aware
robustness by \twitterbaselinetimes{}.
From each cost family, our best models with the strongest
cost-aware robustness are M2, M6, M13, and M19. 
Compared to the natural model obtained from regular training, our robust models
increase the adaptive attack cost to evade them by
10.6$\times$ (M2, $Cost_1$), 10$\times$ (M6, $Cost_2$), 8.3$\times$ (M13, $Cost_3$),
and 6.4$\times$ (M19, $Cost_4$), respectively.
Thus, the highest cost-aware robustness increase is obtained by
M2 model over the total feature manipulation cost $Cost_1$.

\noindent\textbf{Advantages of cost-driven constraint:} Our robust training method using cost-driven 
constraints can achieve stronger cost-aware robustness, higher accuracy,
and lower false positive rate than $L_\infty$-norm cost model from Chen's algorithm~\cite{chen2019robust}.
In Table~\ref{tab:costmodels},
results from C1, C2, and C3 models demonstrate that if we use $L_\infty$-norm cost model ($L_\infty \leq \epsilon$), the performance of the
trained model quickly degrades as $\epsilon$ gets larger.
In particular, when $\epsilon = 0.03$, the C1 model trained by
Chen's algorithm has decreased the accuracy to 96.59\% and
increased the false positive rate to 5.49\% compared to regular training.
With larger $\epsilon$
values, C2 and C3 models have even worse performance. C3
has only 91.89\% accuracy and a very high 11.96\% false positive rate.
In comparison, if we specify attack cost-driven constraint in our
training process, we can train cost-aware robust models with better performance.
For example, our model M6 can achieve stronger robustness against cost-aware attackers
with all four adaptive attack cost than C1. At the same time,
M6 has higher accuracy and lower false positive rate than C1.

\noindent\textbf{Robustness and accuracy tradeoffs:} Training a larger bounding box
generally decreases accuracy and increases false positive rate within the same
cost family; however, the obtained robustness against MILP attacks vary across
different cost families. Whenever we specify a new cost family with
different constraint variable proportions and number of categories,
we need to perform constraint parameter tuning to find the model that best
balances accuracy and robustness.
\begin{itemize}[leftmargin=*]
\setlength\itemsep{0em}
	\item In the first cost family, as the bounding box size increases, the adaptive evasion cost $Cost_1$ against the models increases, and then decreases. M2 has the largest evasion cost.
	\item In the second cost family where we merge medium cost and high cost, the adaptive
	evasion cost $Cost_2$ decreases as the bounding box size increases.
	\item In the third cost family where we merge negligible cost and low cost, the adaptive evasion cost $Cost_3$ has high values for M10 and M13, and varies for
	other models.
	\item In the last cost family where we merge low cost and medium cost, the adaptive evasion cost $Cost_4$ increases as the bounding box size increases.
\end{itemize}

\noindent\textbf{Other mathematical distances:} Although our current implementation does not
support training L1 and L2 attack cost models directly, training our proposed cost models
can obtain robustness against L1 and L2 attacks. Comparing to the C1 model trained by Chen's algorithm,
we can obtain stronger robustness against L1 and L2 based MILP attacks while achieving
better model performance. For example, our models M6 and M19 have larger L1/L2 evasion
distance than C1, and they have lower false positive rate and higher/similar accuracy.

\subsubsection{Discussion}
\label{sec:discussion}

\noindent\textbf{Robustness and accuracy tradeoffs.}
Obtaining robustness of a classifier naturally comes with the tradeoff
of decreased accuracy and increased false positive rate.
We have experimented with 19 different cost models to demonstrate
such tradeoffs in Table~\ref{tab:costmodels}.
In general, we need to perform constraint hyperparamters tuning
to find the model that best balances accuracy, false positive rate, and robustness. 
In comparison with $L_\infty$ based cost models (C1, C2 and C3), we can achieve
relatively higher accuracy and lower false positive rate
while obtaining stronger robustness against cost-aware attackers (e.g., M6 vs C1).
This is because $L_\infty$ based cost model allows attackers to perturb
all features with equally large range, making it harder to achieve such robustness
and easier to decrease the model performance. However, our cost-driven training technique
can target the trained ranges according to the semantics of the features.

\noindent\textbf{Scalability.} For applications where thousands of features are used to build
a classifier, we can categorize the features
by semantics, and specify the cost-driven constraint as a function for
different categories. Alternatively, we can also use $L_\infty$-norm
as default perturbation for features, and specify cost-driven constraint for
selected features.

\noindent\textbf{Generalization.}
Our cost-aware training technique can generalize to any decision tree and tree ensemble training process,
for both classification and regression tasks, e.g., AdaBoost~\cite{friedman2000additive}
and Gradient Boosting Machine~\cite{friedman2001greedy}.
Since we apply the cost-aware constraint in the node splitting process,
when constructing the classification and regression trees, we can calculate the maximal error
of the split construction according to the allowable perturbations of the training data,
and adjust the score for the split. This can be integrated in many different tree ensemble training
algorithms. We leave investigation of integrating our technique to other datasets as future work.

\section{Conclusion}

In this paper, we have designed, implemented, and evaluated a cost-aware robust training method
to train tree ensembles for security.
We have proposed a cost modeling method to capture the domain knowledge about
feature manipulation cost, and a robust training algorithm to integrate such knowledge.
We have evaluated over four benchmark datasets against the regular training method and the state-of-the-art
robust training algorithm. Our results show that compared to the state-of-the-art robust training algorithm,
our model is \gbdtrobusttimes{} more robust in gradient boosted decision trees, and
\sklearnrfrobusttimes{} more robust in random forest models, against
the strongest whitebox attack based on $L_p$ norm.
Using our method, we have trained cost-aware robust Twitter spam detection models
to compare different cost-driven constraints. Moreover,
one of our best robust models can increase the robustness by \twitterbaselinetimes{} against the adaptive attacker.


\section*{Acknowledgements}

We thank Huan Zhang and the anonymous reviewers for their constructive and valuable feedback. This work is supported in part by NSF grants CNS-18-42456, CNS-18-01426, CNS-16-17670, CNS-16-18771, CCF-16-19123, CCF-18-22965, CNS-19-46068; ONR grant N00014-17-1-2010; an ARL Young Investigator (YIP) award; a NSF CAREER award; a Google Faculty Fellowship; a Capital One Research Grant; a J.P. Morgan Faculty Award; and Institute of Information \& communications Technology Planning \& Evaluation (IITP) grant funded by the Korea government(MSIT) (No.2020-0-00153). Any opinions, findings, conclusions, or recommendations expressed herein are those of the authors, and do not necessarily reflect those of the US Government, ONR, ARL, NSF, Google, Capital One, J.P. Morgan, or the Korea government.

\small
\bibliographystyle{abbrv}
\bibliography{ref}

\appendix
\normalsize
\balance

\section{Appendix}

\subsection{Hyperparameters for GBDT Models}
\label{appendix:params_gbdt}

To evaluate the hyperparameters for gradient boosted decision tree models used in~\cite{chen2019robust},
we train 35 models for each dataset to conduct a grid search.
In details, we use maximum depth: 4, 5, 6, 7, 8, 9, 10; number of trees
for breast-cancer: 2, 4, 6, 8, 10, cod-rna: 10, 20, 30, 40, ijcnn1: 20, 40, 60, 80, and binary mnist: 600, 800, 1000, 1200.
Table~\ref{tab:gbdt_gridparam} and~\ref{tab:gbdt_gridsearch} reports the model hyperparameters and corresponding test accuracy of trained models which obtain the best validation accuracy.
In comparison with the results from Table~\ref{tab:gbdt}, the hyperparameters used by~\cite{chen2019robust}
can train models with accuracy similar to the best one.

\begin{table}[!hbt]
	\centering
	\tabcolsep=2.0pt
	\small
	\begin{tabular}{|c|cc|ccc|}
		\hline
		\multirow{2}{*}{Dataset} & \multicolumn{2}{c|}{Trained $\epsilon$} & \multicolumn{3}{c|}{Tree Num / Depth} \\
		& Chen's & ours & natural & Chen's & ours  \\\hline
		breast-cancer & 0.30 & 0.30 & 4 / 6 & 4 / 4 & 2 / 7 \\\hline
		cod-rna & 0.20 & 0.03 & 40 / 10 & 40 / 4 & 10 / 10  \\\hline
		ijcnn1 & 0.20 & 0.02 & 80 / 5 & 80 / 10 & 80 / 10  \\\hline
		MNIST 2 vs. 6 & 0.30 & 0.30 & 600 / 4 & 600 / 8 & 600 / 9 \\\hline
	\end{tabular}
	\caption{GBDT model hyperparameters with the best validation accuracy in XGBoost.}
	\label{tab:gbdt_gridparam}
	\vspace{-10pt}
\end{table}

\begin{table}[!hbt]
	\centering
	\tabcolsep=2.0pt
	\small
	\begin{tabular}{|c|ccc|ccc|}
		\hline
		\multirow{2}{*}{Dataset} & \multicolumn{3}{c|}{Test ACC (\%)} & \multicolumn{3}{c|}{Test FPR (\%)} \\
		& natural & Chen's & ours  & natural & Chen's & ours \\\hline
		breast-cancer & 97.81 & 96.35 & 99.27 & 0.98 & 0.98 & 0.98 \\\hline
		cod-rna & 96.74 & 87.32 & 91.08 & 2.79 & 4.05 & 8.71 \\\hline
		ijcnn1  &  97.85 & 97.24 & 93.66 & 1.74 & 1.53 & 1.70 \\\hline
		MNIST 2 vs. 6 & 99.70 & 99.65 & 99.55 & 0.39 & 0.39 & 0.29 \\\hline
	\end{tabular}
	
	\caption{Test accuracy of GBDT models with the best validation accuracy in XGBoost.}
	\label{tab:gbdt_gridsearch}
\end{table}

\subsection{Recall for Twitter Spam Models}
\label{appendix:recall}

To evaluate the performance of all 23 models trained to detect Twitter spam,
we computed the recall at 1\% FPR, 5\% FPR, and 10\% FPR in Table~\ref{tab:twittermodels}.
The models M1, M6, M10, and M16 have the best recall within their cost family.

\begin{table}[ht!]
	\centering
	\small
	\tabcolsep=5.8pt
	\begin{tabular}{c | c | crc }
		\hline
		\multirow{3}{*}{\begin{tabular}[l]{@{}l@{}}{\textbf{Classifier}}\\{\textbf{Model}}\end{tabular}} & \multirow{3}{*}{\begin{tabular}[l]{@{}l@{}}{\textbf{Adaptive}}\\{\textbf{Objective}}\end{tabular}} & \multicolumn{3}{c}{\multirow{1}{*}{\textbf{Model Quality}}} \\
		& & 1\% FPR & 5\% FPR & 10\% FPR \\
		& & Recall & Recall & Recall \\
		\hline
		\hline
		Natural & - & 0.9974 &  0.9998 &  0.9999 \\ \hline
		C1 & - &  0.8177 &  0.9844 &  0.9999 \\ \hline
		C2 & - & 0.7912 &  0.9250 &  0.9897 \\ \hline
		C3 & - & 0.6928 &  0.8609 &  0.8609 \\ \hline\hline
		M1 & \multirow{5}{*}{$Cost_1$} & \textbf{0.9612} &  \textbf{0.9992} &  \textbf{0.9997} \\ \cline{3-5}
		M2 & & 0.7949 &  0.9893 &  0.9973 \\ \cline{3-5}
		M3 & & 0.8214 &  0.9948 &  0.9981 \\ \cline{3-5}
		M4 & & 0.7537 &  0.9281 &  0.9689 \\ \cline{3-5}
		M5 & & 0.6907 &  0.9280 &  0.9840 \\ \hline
		M6 & \multirow{4}{*}{$Cost_2$} & \textbf{0.9162} &  \textbf{0.9948} & \textbf{0.9968} \\ \cline{3-5}
		M7 &  &  0.7881 &  0.9901 &  0.9959 \\ \cline{3-5}
		M8 &  &  0.6793 &  0.9220 &  0.9608 \\ \cline{3-5}
		M9 &  &  0.6780 &  0.9016 &  0.9386 \\ \hline
		M10 & \multirow{6}{*}{$Cost_3$} & \textbf{0.9715} & \textbf{0.9996} &  \textbf{0.9999} \\ \cline{3-5}
		M11 &  &  0.8671 &  0.9948 &  0.9991 \\ \cline{3-5}
		M12 &  &  0.7484 &  0.9846 &  0.9930 \\ \cline{3-5}
		M13 &  &  0.7753 &  0.9383 &  0.9896 \\ \cline{3-5}
		M14 &  &  0.7473 &  0.9806 &  0.9925 \\ \cline{3-5}
		M15 &  &  0.6728 &  0.8852 &  0.9862 \\ \hline
		M16 & \multirow{4}{*}{$Cost_4$} & 0.8624 &  0.9929 & \textbf{0.9989} \\ \cline{3-5}
		M17 &  &  \textbf{0.9061} &  \textbf{0.9946} &  0.9973 \\ \cline{3-5}
		M18 &  &  0.7075 &  0.9368 &  0.9749 \\ \cline{3-5}
		M19 &  &  0.7298 &  0.9361 &  0.9703 \\ \hline
	\end{tabular}
	\caption{
		Recall at 1\% FPR, 5\% FPR, and 10\% FPR for all Twitter spam detection models.
		The best recall numbers highlighted in bold.}
	\label{tab:twittermodels}
\end{table}

%
%

\end{document}